\documentclass[11pt]{article}

\usepackage[letterpaper,margin=1in]{geometry}
\usepackage[T1]{fontenc}
\usepackage[utf8]{inputenc}
\usepackage{lmodern}
\usepackage{microtype}

\usepackage{amsmath,amssymb}
\usepackage{booktabs,array,changepage,longtable}
\usepackage[table]{xcolor}
\usepackage{graphicx}

\usepackage[colorlinks=true,linkcolor=blue!45!black,urlcolor=blue!45!black,%
            citecolor=blue!45!black]{hyperref}
\usepackage{textcomp}

\usepackage{authblk}
\usepackage{orcidlink}

\usepackage[labelfont=bf,labelsep=period,justification=raggedright,%
            singlelinecheck=off,font=small]{caption}

\usepackage{float}     
\usepackage{flafter}
\usepackage{placeins}

\usepackage{cite}

\usepackage{fancyhdr}
\title{\bfseries\LARGE Cross-cultural evaluation of taste--sound\\ correspondences in AI-generated music}

\author[1]{Matteo Spanio\,\orcidlink{0000-0002-2436-7208}\thanks{Corresponding author: \texttt{spanio@dei.unipd.it}}}
\author[2]{Massimiliano Zampini\,\orcidlink{0000-0001-5950-7365}}
\author[3]{Luisa Torri\,\orcidlink{0000-0003-4522-6452}}
\author[3]{Riccardo Migliavada\,\orcidlink{0000-0002-4262-8360}}
\author[4]{Bruno Mesz\,\orcidlink{0000-0002-4941-818X}}
\author[5]{Masaki Ohno\,\orcidlink{0000-0002-8404-112X}}
\author[5]{Yuji Wada\,\orcidlink{0000-0002-4982-6562}}
\author[1]{Antonio Rod\`a\,\orcidlink{0000-0001-9921-0590}}

\affil[1]{University of Padova, Padova, Italy}
\affil[2]{University of Trento, Trento, Italy}
\affil[3]{University of Gastronomic Sciences, Pollenzo, Italy}
\affil[4]{Universidad Nacional de Tres de Febrero, Buenos Aires, Argentina}
\affil[5]{Ritsumeikan University, Kyoto, Japan}

\date{\vspace{-2em}}

\begin{document}
\maketitle
\thispagestyle{fancy}

\begin{abstract}
\noindent
Sonic seasoning research has shown that listeners attribute systematic gustatory and emotional
meaning to sound, and text-to-music generative artificial intelligence (AI) has recently been used to
render gustatory prompts as musical stimuli. Whether the taste--sound correspondences acquired by such
models hold beyond the cultural context in which they were validated remains untested.
We extended a single-country Italian study to a three-country online experiment conducted in
Argentina, Italy, and Japan ($N = 361$), cohorts selected to span three continents and three distinct
culinary and musical traditions. Participants first indicated their preference between base and
fine-tuned MusicGen excerpts generated from four taste prompts (sweet, sour, bitter, salty), and then
rated fine-tuned excerpts on twelve taste, emotion, and thermal descriptors.
Preference for the fine-tuned model was confirmed in Argentina and Italy but not in Japan, and the
salty prompt yielded the weakest correspondence in all three cohorts. Ratings differed substantially
between countries, yet the main effect of country was no longer detectable once ratings had been
standardized within participant, whereas the interactions characterizing the mapping of prompts onto
descriptors remained essentially unchanged. Much of the apparent cross-cultural divergence is
therefore attributable to differences in scale use; a structural component nevertheless persists.
Excerpts occupying the same acoustic region, characterized by high spectral roughness and sensory
dissonance, were predominantly labelled sour in Italy and Japan but bitter in Argentina, and
exploratory factor analysis indicated that the twelve descriptors were organized along different
latent dimensions in each cohort.
These results indicate that cross-cultural variation in AI-mediated sonic seasoning operates at two
levels: the overall level at which taste is attributed to a given stimulus, and the relational
structure of those attributions. Evaluations of generative music systems across populations should
accordingly distinguish response-style bias from genuine perceptual reorganization.

\end{abstract}

\vspace{0.5em}
\noindent\textbf{Keywords:} crossmodal correspondences \textperiodcentered\ sonic seasoning
\textperiodcentered\ taste--sound associations \textperiodcentered\ generative AI
\textperiodcentered\ text-to-music \textperiodcentered\ cross-cultural perception
\textperiodcentered\ response style

\vspace{1em}
\hrule
\vspace{1em}

\graphicspath{{figures/}}

\section{Introduction}

Crossmodal correspondences refer to the systematic associations that people consistently make between features belonging to different sensory modalities, such as linking high-pitched sounds with sweetness or low-pitched sounds with bitterness, despite the absence of any direct physical relationship between them \cite{Spence2021SonicSeasoningCoffee,ReinosoCarvalho2020Blending}. These correspondences are thought to influence perceptual expectations and multisensory integration, providing a theoretical basis for understanding how information from one sensory modality can shape experience in another \cite{Spence2021SonicSeasoningCoffee}. In this context, sonic seasoning refers to the deliberate use of music or soundscapes to enhance or modify taste and flavor perception by exploiting these auditory--gustatory correspondences \cite{Spence2021SonicSeasoningCoffee,ReinosoCarvalho2020Blending,Guedes2024Disentangling}. Although growing evidence suggests that sonic seasoning can reliably bias taste perception, recent work has also shown that these effects are partly mediated by the emotional qualities conveyed by music, highlighting the need to disentangle the respective contributions of crossmodal correspondences and affect \cite{ReinosoCarvalho2020Blending,Guedes2024Disentangling}.

At the same time, the apparent robustness of many crossmodal correspondences coexists with growing evidence that their expression is shaped by cultural context \cite{spence_group_2022,wan_crosscultural_2014,velasco_tasteshape_2016, di_stefano_consonance_2026}. Rather than being entirely universal or entirely culture-specific, these correspondences arise from a combination of structural, statistical, semantic, and affective mechanisms, some of which are more susceptible to cultural learning than others \cite{spence_group_2022}. Cross-cultural studies have shown that, although several musical and taste-related correspondences are consistently observed across populations, others vary according to cultural background, reflecting differences in sensory experience, language, and learned associations rather than a simple East--West distinction \cite{wan_crosscultural_2014,velasco_tasteshape_2016}. This issue is particularly relevant in the musical domain, where auditory--conceptual associations are shaped not only by structural regularities but also by culturally acquired affective meanings \cite{di_stefano_prokofiev_2024}. Similarly, familiarity with a musical system and long-term enculturation influence both emotional interpretation and the semantic organization of musical stimuli across cultures \cite{li_cross_cultural_biases_2025,balkwill_japanese_2004}. Consequently, identical musical stimuli may evoke different semantic and affective responses across cultures, potentially leading to different auditory--gustatory correspondences \cite{pengli_crosscultural_2020}.

These developments have recently been extended to generative artificial intelligence (AI), where text-to-music models are used to translate gustatory prompts into musical stimuli and subsequently evaluate whether listeners perceive the intended sensory qualities in the generated outputs \cite{spanio_frontiers_2025,deng_composing_palate_2026}. Rather than functioning solely as creative systems, generative models have begun to serve as experimental tools for investigating whether computational models can learn and reproduce human auditory--gustatory correspondences and, generally speaking, cross-modal patterns \cite{spanio2024aixia}. Spanio et al.~\cite{spanio_frontiers_2025} fine-tuned a MusicGEN model and evaluated its outputs through two complementary experiments: a model-comparison task assessing whether fine-tuning improved taste coherence and a semantic-rating task measuring gustatory, emotional, and thermal responses elicited by the generated music. Using a different approach, Deng et al.~\cite{deng_composing_palate_2026} combined professional sound designers with a human-in-the-loop text-to-audio generation pipeline to create taste-congruent soundscapes. Although the two studies adopted substantially different generation strategies and evaluation procedures, participants in both were able to reliably recover taste-related information from AI-generated auditory stimuli. Together, these findings demonstrate that generative AI provides a controlled and reproducible experimental platform for investigating auditory--gustatory correspondences. In particular, the methodology introduced by Spanio et al. makes it possible to generate standardized musical stimuli from identical gustatory prompts and evaluate them systematically using shared semantic scales across different groups of listeners.

Despite these promising findings, the extent to which AI-mediated auditory--gustatory correspondences generalize across cultures remains unknown. Although the fine-tuned MusicGEN model outperformed its non-fine-tuned counterpart in taste-matching judgments \cite{spanio_frontiers_2025}, the original validation relied on a relatively homogeneous participant sample and therefore could not determine whether the learned correspondences extend beyond the cultural context in which they were acquired. This limitation is reinforced by recent work on culturally adaptive music representation learning, which argues that current foundation models remain strongly influenced by predominantly Western training data and cannot be assumed to transfer uniformly across musical traditions \cite{kanatas_culturemert_2025}. Consequently, if generative models learn auditory--gustatory correspondences from culturally situated data, it remains unclear whether they capture broadly shared perceptual regularities or instead reproduce culturally specific semantic and affective associations.

A second challenge concerns measurement. Cross-cultural comparisons based on semantic rating scales may reflect not only genuine perceptual differences but also systematic differences in response style, including acquiescence and extreme responding \cite{harzing_response_2006,fischer_standardization_2004,he_bias_2013}. Interpreting cross-cultural variation therefore requires distinguishing differences in perception from differences in the way participants use rating scales.

This study investigates whether AI-mediated sonic seasoning generalizes across culturally distinct populations and, if not, what accounts for the observed differences. Specifically, we examine whether the superiority of a fine-tuned model over its non-fine-tuned counterpart can be replicated beyond the cultural context in which it was originally validated, whether listeners from different countries organize auditory--gustatory correspondences in similar ways, and whether any observed cross-cultural variation reflects genuine differences in perceptual organization, culturally specific semantic associations, response styles, or a combination of these factors.

To address these questions, we conducted a cross-cultural online experiment involving participants from Argentina, Italy, and Japan. These countries were selected to capture diversity in both culinary and musical traditions while maintaining a balanced three-country comparison. All three have well-established culinary traditions in which food plays an important cultural and social role. Italian cooking and Japanese washoku are recognized by UNESCO as elements of the Intangible Cultural Heritage of Humanity \cite{unesco_italian_cooking_2025,unesco_washoku_2013}, while studies of Argentine gastronomy likewise identify food as a central component of cultural identity and social life \cite{cuffia_argentina_gastronomy_2023}. Because the experimental stimuli consist of taste prompts, cultural differences in the representation and interpretation of basic tastes are directly relevant to the way listeners may associate taste with sound.

The three countries also differ substantially in their musical traditions. Italy serves as the reference condition because the present study directly extends the original validation conducted there \cite{spanio_frontiers_2025}. Japan provides a culturally distant comparison in which musical enculturation has been shown to influence the perception of emotional meaning in music \cite{balkwill_japanese_2004}. Argentina contributes a distinct Western perspective, combining a rich musical tradition with an established body of research on taste--music correspondences \cite{mesz_taste_music_2011}. Including Argentina also makes it possible to distinguish differences associated with broader cultural distance from those that may emerge between culturally distinct Western populations, avoiding a comparison restricted to Europe and East Asia alone.

\begin{figure}[!ht]
\centering
\includegraphics[width=\linewidth]{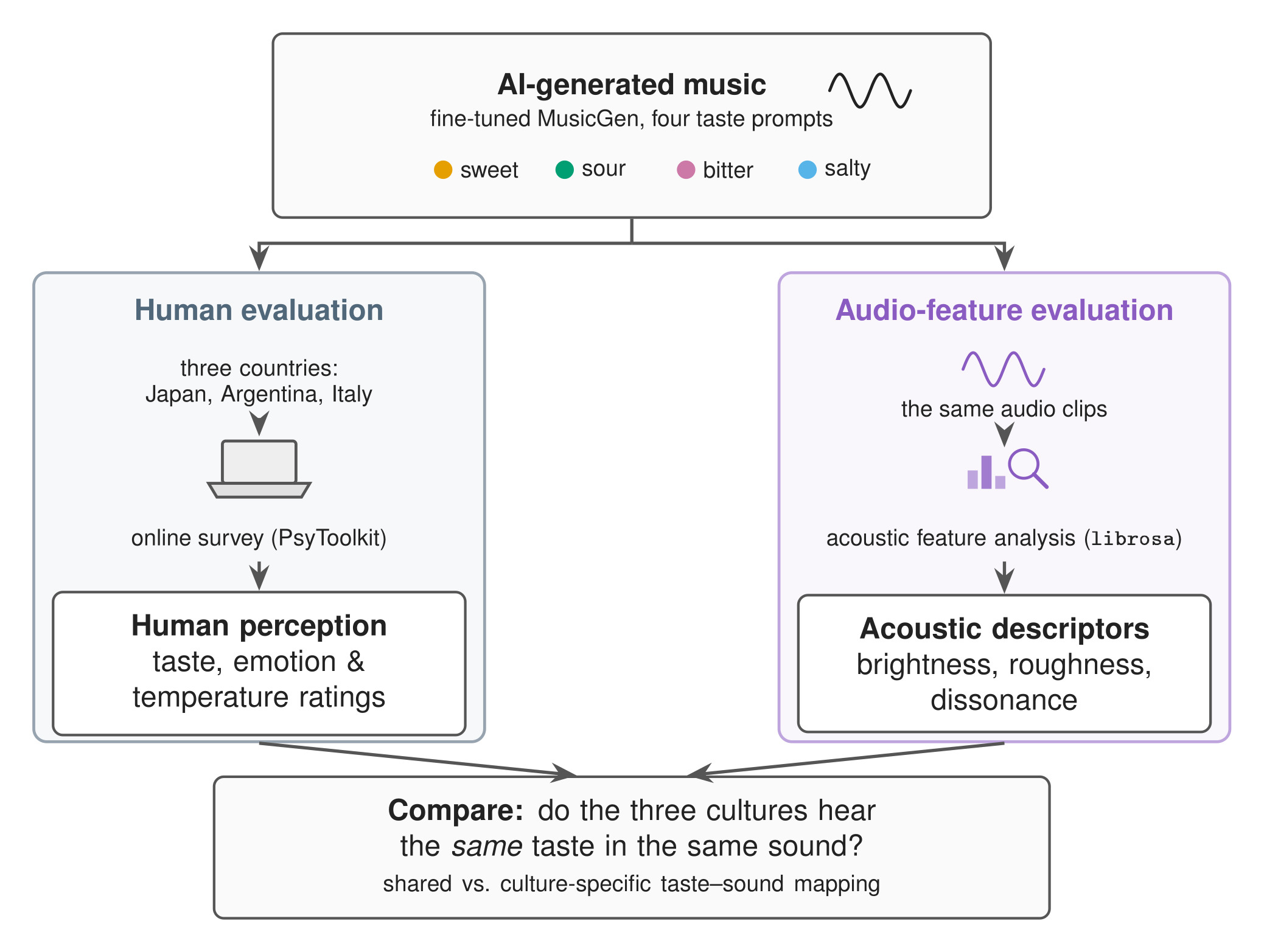}
\caption{Conceptual overview of the study. AI-generated music---fine-tuned MusicGen clips for four
taste prompts (sweet, sour, bitter, salty)---is evaluated in two parallel ways and the results are
compared. Human listeners in three countries (Japan, Argentina, and Italy) rate the clips through an
online survey, yielding subjective taste, emotion, and thermal judgments, while an acoustic
feature analysis (\texttt{librosa}) characterizes the same clips on descriptors such as brightness,
roughness, and dissonance. Comparing the two asks whether listeners from different cultures hear the
same taste in the same sound.}
\label{fig:overview}
\end{figure}

Taken together, this design allows us to evaluate whether AI-mediated auditory--gustatory correspondences remain stable across culturally diverse populations while separating three sources of variation that are often conflated in cross-cultural studies: overall preference for the generated stimuli, culturally specific response styles, and differences in the organization of auditory--gustatory correspondences themselves. Figure~\ref{fig:overview} provides an overview of the study design.

\FloatBarrier

\section{Materials and methods}

\subsection{Participants}
\label{participants}

The study included three national cohorts recruited independently in Italy, Japan, and Argentina. The Italian cohort combined respondents from the original Italian dataset of Spanio et al.~\cite{spanio_frontiers_2025}, restricted to participants who self-identified as ethnically Italian, with an additional Italian cohort recruited subsequently using the same questionnaire, auditory stimuli, and experimental protocol. The Japanese and Argentine cohorts were recruited specifically for the present study. Because the previously collected and newly recruited Italian participants completed an identical experimental procedure, their responses were pooled for all subsequent analyses. After excluding incomplete questionnaires, the final sample comprised $N = 361$ participants (Italy: $n = 117$, Japan: $n = 140$, Argentina: $n = 104$). An additional English-language cohort was also collected but was excluded from the present analyses because its culturally heterogeneous composition was not compatible with the three-country comparison.
The study protocol and questionnaire were approved by the Ethics Committee of the University of Gastronomic Sciences of Pollenzo (UNISG Ethics Committee, Minutes 19122025). The study was conducted in accordance with the ethical principles of the Declaration of Helsinki and its later amendments (October 2024). Participation was voluntary and restricted to adults (18 years or older). Before beginning the survey, all participants provided informed consent electronically through PsyToolkit \cite{stoet_psytoolkit_2017}. No personally identifying information was collected, and all analyses were performed on anonymized data

\subsection{Materials and stimuli}

The present study employed the same auditory stimuli and experimental materials used by Spanio et al.~\cite{spanio_frontiers_2025}, allowing a direct cross-cultural replication of the original Italian experiment. The stimulus set consisted of short musical excerpts generated with the MusicGen text-to-music model \cite{copet_musicgen_2024} using two versions of the model: the original public model and the fine-tuned model proposed by Spanio et al. Each model generated four musical excerpts corresponding to the taste prompts \emph{sweet}, \emph{sour}, \emph{bitter}, and \emph{salty}. All prompts were specified in English, the language of MusicGen's T5 text encoder.
No modifications were made to the original stimuli or experimental protocol. Consequently, responses from the original Italian dataset were directly comparable with those collected from the newly recruited Italian, Japanese, and Argentine participants, allowing all observations to be analyzed within a common experimental framework (see Section~\ref{participants}).
For each participant, the assignment of prompts to audio clips, the presentation order of the clips within each task, and the order of the experimental trials were independently randomized.

\subsection{Procedure}

The experiment was administered online using PsyToolkit \cite{stoet_psytoolkit_2017}. The questionnaire was available in four languages (Italian, Japanese, Spanish, and English), and participants completed the survey in their native language whenever possible. The English version was administered exclusively to the international cohort, which was excluded from the present analyses (see Section~\ref{participants}). Apart from translation, all versions of the questionnaire were identical in structure, stimuli, response scales, and experimental flow.
The survey comprised a demographic questionnaire followed by two listening tasks. The first task (\emph{Model Preference}) assessed participants' preference between musical excerpts generated by the original and fine-tuned MusicGen models: on each of five trials participants heard one clip from each model, both generated from the same taste prompt, and placed their preference on a 0--10 slider anchored at the two clips, with 5 as the neutral midpoint. Responses are reported throughout so that values above 5 indicate preference for the fine-tuned model. The second task (\emph{Semantic Rating}) required participants to evaluate the same musical excerpts on twelve 1--5 rating scales covering gustatory, emotional, and thermal dimensions. We refer to these twelve rating items collectively as \emph{descriptors}: four \emph{taste descriptors} (sweet, sour, bitter, salty), six emotion descriptors (happy, sad, anger, disgust, fear, surprise), and two thermal descriptors (hot, cold). Throughout the paper, \emph{prompt} is reserved for the taste used to condition the generative model, so that the input given to MusicGen and the descriptor rated by participants are never confused even when the two coincide. Prompt-to-audio assignment, clip order within each task, and task order were independently randomized for every participant.
Sample size was determined through the Monte Carlo power analysis reported in Appendix~\ref{app:si}. Based on the preliminary Italian dataset, approximately 80 participants per country were estimated to provide 80\% statistical power to detect the expected medium-sized effect of prompt. All three cohorts exceeded this target.
The demographic questionnaire collected participants' age, gender, self-reported ethnicity, country of origin, country of residence, self-reported musical expertise (\emph{hearing experience}), and self-reported food expertise (\emph{eating experience}). Country names entered as free text were harmonized to standardized English labels before analysis.

\subsection{Statistical analysis}

All analyses were carried out in R \cite{r_core_2025}. The complete analysis code (data loading,
models, figures, and tables) is openly available at
\url{https://github.com/matteospanio/multimodal-survey-analysis}, and the survey data and audio stimuli
are archived on Zenodo (DOI: \href{https://doi.org/10.5281/zenodo.20841611}{10.5281/zenodo.20841611}).

\subsubsection{Demographic equivalence}

We first tested whether the independently recruited cohorts were comparable on background variables. Age
was compared with a Kruskal--Wallis test; gender, hearing experience, and eating experience with Fisher's
exact tests on the contingency tables (Monte Carlo simulation for sparse cells). Variables that differed
across groups were carried into the perception-rating models as nuisance covariates.

\subsubsection{Task 1: model preference}

Because the five trials are nested within participants, the unit of analysis is the participant: we
averaged each participant's trials (overall and within each taste prompt shown) and ran all inferential
tests on those means. The participant means were non-normal (Shapiro--Wilk $p < .001$), so the tests
were non-parametric. Within each country we tested $H_0: \mu = 5$ against $H_1: \mu > 5$ with a Wilcoxon
signed-rank test, where values above 5 indicate preference for the fine-tuned model, reporting the
rank-biserial correlation $r$ (a non-parametric effect size) with a bootstrap $95\%$ confidence interval.
Between-country differences used a Kruskal--Wallis $H$ test (pooled and per prompt; the per-prompt tests
in Appendix~\ref{app:si} (Table~A5) are uncorrected), reporting the $H$-based $\eta^2$; significant omnibus
effects were followed by Dunn's pairwise comparisons with Bonferroni correction \cite{dunn_multiple_1964}.
Within-country per-prompt Wilcoxon tests (Bonferroni-corrected) identified which prompts drove the pooled
preference. As a robustness check on the trial-level data, we additionally fitted linear mixed models
with random intercepts for participant and stimulus, both within each country (against the neutral point)
and across countries (fixed effect of country).

\subsubsection{Task 2: perception ratings}

For the rating task, the 1--5 value was the dependent variable, with data in long format (one
(participant, prompt, descriptor) triple per observation). Following \cite{spanio_frontiers_2025}, we
restricted the model to respondents reporting Male or Female gender and not classified as
\emph{Professional Eaters} (too sparse for reliable inference).

The main fixed-effects model was:
\begin{equation*}
\text{value} \sim \text{group} \times \text{prompt} \times \text{descriptor}
              + \text{hearing\_experience} + \text{eating\_experience} + \text{gender} .
\end{equation*}
We estimated Type~III sums of squares with \texttt{car::Anova}, using sum-to-zero contrasts so that
lower-order terms remain interpretable under the three-way interaction. Residual heteroscedasticity was
present but tolerable at this sample size \cite{glass_assumptions_1972, harwell_montecarlo_1992,
lix_variance_1996}. The full ANOVA tables, including sums of squares, are given in
Appendix~\ref{app:si} (Tables~A7 and~A8); we also ran per-country ANOVAs
without the \emph{group} factor (Table~A9).

Because each participant contributed 36 nested ratings and several participants rated the same excerpt,
we also fitted a linear mixed model with the same fixed-effects structure and crossed random intercepts
for participant and stimulus song, $(1 \mid \text{participant}) + (1 \mid \text{song})$, as a
participant- and stimulus-level robustness check. We retain the fixed-effects ANOVA as the main table for
comparability with the earlier Italian study but treat an ANOVA result as supported only when the mixed
model agrees.

We further checked the cross-cultural conclusion by re-fitting the omnibus model and comparing the four
group-related terms (i) retaining the respondents excluded above (gender other than Male or Female, and
Professional Eaters) and (ii) excluding the legacy Italian wave, and by fitting an Italian-only mixed
model testing collection wave and its interaction with the prompt~$\times$~descriptor structure.

\subsubsection{Response-style diagnostics and sensitivity analysis}

Response styles---using the scale in a particular way independently of item content---can confound
cross-cultural ratings \cite{harzing_response_2006, fischer_standardization_2004, he_bias_2013}, so we
computed two participant-level diagnostics: the Acquiescence Response Style index (ARS), the mean of each
participant's 36 ratings, and the Extreme Response Style index (ERS), the proportion of those ratings
equal to 1 or 5. Both were compared across groups with Kruskal--Wallis tests; the indices and a
complementary variance-homogeneity (Levene) check are tabulated in
Appendix~\ref{app:si} (Tables~A13--A15).

We then re-fitted the omnibus model on within-subject z-scored ratings (each rating minus the
participant's mean over all 36 ratings, divided by the within-participant standard deviation; one
participant with zero within-subject variance was excluded here only). Because this normalization removes
between-participant location and scale by design, the informative terms are
\emph{group} $\times$ \emph{descriptor} and \emph{group} $\times$ \emph{prompt} $\times$ \emph{descriptor},
which capture changes in the structure of the mapping rather than in overall scale use.

\subsubsection{Inter-annotator agreement}

For each (song, taste-descriptor) cell we treated its raters as annotators and computed two agreement
indices on the 1--5 Likert rescaled to $[0, 1]$: the mean pairwise distance (MPD), the average
$|x_i - x_j|$ between annotators (smaller is better, $0$ indicating perfect agreement), and Krippendorff's
$\alpha$ with the interval metric $\delta^2(a, b) = (a - b)^2$. Because $\alpha$ is invariant to a common
linear transform, it is identical on raw or within-country z-scored ratings; we also computed an
ordinal-metric variant as a sensitivity check. Per cohort, $\alpha$ was aggregated over the four taste
descriptors by pooling the matching song--descriptor cells (100 songs, 25 per prompt; cells with fewer
than two raters dropped); the per-cohort values are tabulated in Appendix~\ref{app:si}
(Table~A16).

\subsubsection{Factor structure comparison}

The twelve descriptor ratings were submitted to exploratory factor analysis (maximum-likelihood
extraction, oblimin rotation) with the \texttt{psych} package \cite{revelle_psych_2025}, the number of
factors chosen by parallel analysis per cohort and for the pooled sample (reported in
Appendix~\ref{app:si}, Figs~A4--A6).
To compare structures across countries, we computed Tucker's congruence coefficients
\cite{tucker_congruence_1951} between all pairs of group-specific solutions, truncating each comparison to the smaller number of factors retained; values above $0.95$ indicate factor equivalence and values in $[0.85, 0.95)$ fair similarity. We treated the solutions as exploratory: the Italian three-factor solution returned a Heywood case for the \emph{sweet} item (loading near $1.00$, residual variance near zero), so the Italian structure is a descriptive summary of how the descriptor space partitions in that cohort, not a stable measurement model.

\subsubsection{Acoustic characterization}
\label{sec:acoustic-method}

To test whether low-level acoustic cues underlie the cross-cultural effects, we characterized the 100
fine-tuned clips (25 per prompt) with the eight descriptors used by Deng et al.---spectral centroid
(brightness), bandwidth (frequency content), rolloff (sharpness), fundamental frequency (pitch), spectral flux (roughness),
onset strength (articulation, attack salience), tempo (speed), and zero-crossing rate (discontinuity, noisiness)---computed with
\texttt{librosa} \cite{mcfee_librosa_2015}, plus a Plomp--Levelt sensory-dissonance index in the
parametrization of Sethares \cite{sethares_dissonance_1993}. Following Deng et al., we read these as relative patterns across prompts rather than absolute thresholds.

\section{Results}

The analyses below follow a chain of increasingly specific questions. We first verify that the three
cohorts are comparable on background variables, and then ask whether the fine-tuned model is preferred
over the base model in each country (Task~1). Turning to the rating task (Task~2), we ask whether the
three cohorts describe the same clips in the same way, and whether any country difference reflects how
participants use the rating scale or how they organize the sound-to-taste mapping itself. We then
characterize that organization: how much listeners within a cohort agree with one another, and how the
twelve descriptors group into a smaller number of latent dimensions. Finally, we turn from the listeners
to the stimuli and ask what the generated clips actually sound like, and which taste descriptor each
cohort attaches to a given acoustic profile. Throughout, \emph{prompt} denotes the taste used to
condition the generative model and \emph{descriptor} one of the twelve items participants rated.

\subsection{Sample composition}

The pooled sample comprises 140 Japanese, 104 Argentine, and 117 Italian respondents
(Table~\ref{tab:demographics}) and differs substantially on background variables: age varied by group
($H(2) = 77.89$, $p < .001$), and Fisher's exact tests rejected homogeneity for gender, hearing
experience, and eating experience (all $p < .001$ via Monte Carlo simulation). The cohorts are therefore
not interchangeable on background characteristics, and these variables are carried as covariates in every
rating model reported below.

\begin{table}[!ht]
\centering
\caption{Demographic summary by cultural group. Age is mean $\pm$ standard deviation; gender is
percentage of the within-group sample (\emph{Other / NA} pools \emph{Other} and \emph{Not specified});
time is the median minutes to complete the survey (median because completion time is right-skewed).}
\label{tab:demographics}
\small
\begin{tabular}{lrrrrrr}
\toprule
Group & $N$ & Age (y) & \% Male & \% Female & \% Other / NA & Time (min) \\
\midrule
Japan     & 140 & $27.2 \pm 13.3$ & 28.6 & 68.6 & 2.9 & 8.0 \\
Argentina & 104 & $42.4 \pm 15.0$ & 49.0 & 50.0 & 1.0 & 8.0 \\
Italy     & 117 & $34.2 \pm 14.7$ & 53.8 & 43.6 & 2.6 & 8.0 \\
\midrule
Pooled    & 361 & $33.9 \pm 15.5$ & 42.7 & 55.1 & 2.2 & 8.0 \\
\bottomrule
\end{tabular}
\end{table}

\subsection{Task 1: model preference}

Task~1 asks the simplest question in the study: presented with two clips generated from the same taste
prompt, do listeners prefer the one produced by the fine-tuned model? Preferences are placed on a 0--10
slider with 5 as the neutral midpoint, so a participant mean above 5 favours the fine-tuned model and a
mean below 5 the base model.

The answer is a clear partial replication. One-sided Wilcoxon signed-rank tests on the participant mean
scores show that the fine-tuned model is preferred in Argentina (median $= 5.6$, rank-biserial
$r = .29$, $95\%$ CI $[.08, .51]$, $p = .006$) and in Italy (median $= 5.2$, $r = .30$, $[.09, .50]$,
$p = .003$), but not in Japan (median $= 5.0$, $r = -.15$, $[-.34, .05]$, $p = .93$). The within-country
effect sizes are small in Argentina and Italy and indistinguishable from zero in Japan. Comparing the
three countries directly, the participant means differ, $H(2) = 11.47$, $p = .003$, $\eta^2 = .03$;
Dunn's post-hoc tests attribute this to the two contrasts involving Japan, which differs from both
Argentina ($p_{\text{adj}} = .006$) and Italy ($p_{\text{adj}} = .006$), whereas Argentina and Italy do
not differ ($p_{\text{adj}} = 1.00$). A linear mixed model on the trial-level data, with random
intercepts for participant and stimulus, reproduces this pattern: the estimated preference above neutral
is positive in Argentina ($+0.43$, $p = .012$) and Italy ($+0.47$, $p = .021$) and null in Japan
($-0.15$, $p = .33$), with a significant country effect ($p = .003$). The conclusion therefore does not
depend on how the nesting of trials within participants is handled.

\begin{figure}[!ht]
\centering
\includegraphics[width=0.94\linewidth]{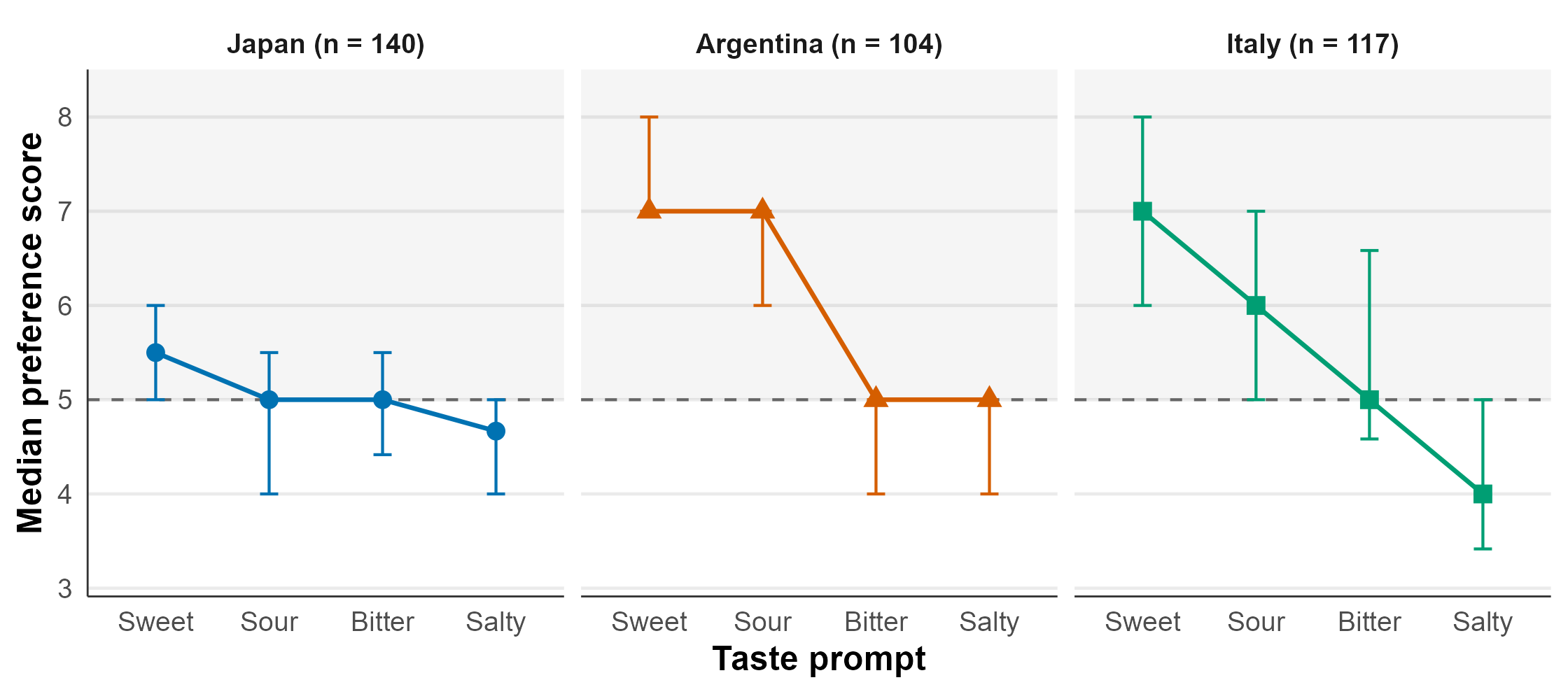}
\caption{Model preference summaries by prompt and cultural group, shown as one panel per cohort.
Points mark the median of the participant mean scores, vertical bars show bootstrap 95\% confidence
intervals, and the dashed horizontal line marks the neutral point of 5. Scores in the shaded band
(above 5) favor the fine-tuned model, those below favor the base model.}
\label{fig:task1-summary}
\end{figure}

Fig~\ref{fig:task1-summary} shows where the advantage comes from. Within Argentina and Italy the
per-prompt breakdown locates the fine-tuning advantage in the sweet and sour prompts
(Bonferroni-corrected $r$ in $[.34, .60]$, all $p_{\text{adj}} \le .010$), and the between-country
difference is concentrated on the same two prompts (sweet $H(2) = 16.45$, $p < .001$; sour
$H(2) = 13.37$, $p = .001$), where Argentine and Italian medians are highest and Japanese medians stay
near neutral. Bitter and salty show no reliable cross-country difference ($p = .094$ and $p = .216$,
uncorrected; full test table in Appendix~\ref{app:si}, Table~A5). Salty is also the weakest matched
prompt overall---at
the neutral point in Argentina ($r = -.16$) and below neutral in Italy ($r = -.46$) and Japan
($r = -.35$). Because it fails in every cohort alike, this points to a limitation of the fine-tuned
model's salty stimuli rather than to a country-specific failure, a reading we return to when we examine
the acoustics of the clips themselves.

\subsection{Task 2: perception ratings}

Task~1 establishes \emph{that} the three cohorts reach different verdicts on the model. Task~2 asks
\emph{how} they hear the clips, by having every participant rate the same fine-tuned excerpts on all
twelve descriptors. The question is whether the three-dimensional pattern of ratings---country by prompt
by descriptor---has the same shape everywhere.

The left half of Table~\ref{tab:anova} reports the omnibus Type~III ANOVA on the raw ratings. The largest
term is the main effect of group ($F(2, 11{,}768) = 436.22$, $p < .001$), that is, a difference in the
overall level at which the three cohorts rate. The critical result, however, is the three-way
\emph{group} $\times$ \emph{prompt} $\times$ \emph{descriptor} interaction
($F(66, 11{,}768) = 2.96$, $p < .001$): this term asks whether the profile of descriptor ratings across
the four prompts has the same shape in all three countries, and it says that it does not. The two-way
\emph{group} $\times$ \emph{descriptor} interaction is also sizeable ($F(22, 11{,}768) = 14.43$,
$p < .001$), while the raw \emph{group} $\times$ \emph{prompt} term is small
($F(6, 11{,}768) = 2.43$, $p = .024$). All of these cross-cultural terms are highly significant but
account for little variance individually (all group-related partial $\eta^2 \le .03$), so they are best
described as structured but modest tendencies rather than wholesale reorganizations of perception.

Two checks confirm that this structure is not an artifact of how the data were modelled or assembled.
First, a linear mixed model with the same fixed-effects structure and crossed random intercepts for
participant and stimulus song, $(1 \mid \text{participant}) + (1 \mid \text{song})$, leaves the three-way
interaction highly significant ($p < .001$) and confirms the \emph{group}, \emph{prompt},
\emph{descriptor}, and \emph{group} $\times$ \emph{descriptor} effects; only the raw
\emph{group} $\times$ \emph{prompt} interaction fails to survive ($p = .428$). We therefore treat the
three-way interaction, not the prompt-level country shift, as the stable cross-cultural signal.
Song-level variance is small relative to the residual ($\hat{\sigma}_{\text{song}} = 0.04$ versus
$\hat{\sigma}_{\text{residual}} = 1.01$), so individual songs are controlled for without driving the
result. Second, retaining the eight respondents with a gender other than Male or Female and the
twenty-two Professional Eaters excluded by the main model, or excluding the legacy Italian wave, leaves
the four group-related terms essentially unchanged, with the \emph{group} $\times$ \emph{descriptor} and
three-way interactions highly significant ($p < .001$) in every variant. Within the Italian cohort,
collection wave has no effect on ratings ($p = .81$) and no interaction with the
prompt~$\times$~descriptor structure ($p = .28$), so the legacy ($n = 90$) and newly collected
($n = 27$) Italian participants are interchangeable here; the legacy-excluded re-fit, resting on only
those $27$, is best read as a directional check rather than a precise re-estimate.

A group main effect of this size is, however, ambiguous. It could mean that Japanese listeners genuinely
heard more taste in the clips, or simply that they used the 1--5 scale differently from the other two
cohorts. The next section separates these two possibilities.

\begin{table}[!ht]
\centering
\caption{Type~III ANOVA of the Task~2 ratings, before (left) and after (right) within-subject
normalization, with sum-to-zero contrasts on all categorical predictors. Numerator $df$ is the same in
both models; denominator $df = 11{,}768$ (raw) and $11{,}732$ (z-scored, one participant with zero
within-subject variance excluded). Reading the two halves side by side isolates what response style
accounts for: the \emph{group} main effect collapses from partial $\eta^2 = .069$ to below $.001$ and
\emph{group} $\times$ \emph{prompt} becomes non-significant, whereas
\emph{group} $\times$ \emph{descriptor} and the three-way interaction are essentially unchanged. Full
tables including sums of squares are given in Appendix~\ref{app:si} (Tables~A7 and~A8).}
\label{tab:anova}
\small
\setlength{\tabcolsep}{4pt}
\begin{tabular}{lrrrrrrr}
\toprule
 & & \multicolumn{3}{c}{Raw ratings} & \multicolumn{3}{c}{Within-subject $z$ scores} \\
\cmidrule(lr){3-5} \cmidrule(lr){6-8}
Term & $df$ & $F$ & $p$ & partial $\eta^2$ & $F$ & $p$ & partial $\eta^2$ \\
\midrule
Group                      & 2  & 436.22 & $< .001$ & .069    & 0.002 & $.998$   & $<.001$ \\
Prompt                     & 3  & 23.28  & $< .001$ & .006    & 11.51 & $< .001$ & .003 \\
Descriptor                 & 11 & 29.64  & $< .001$ & .027    & 35.44 & $< .001$ & .032 \\
Hearing experience         & 2  & 31.62  & $< .001$ & .005    & 0.03  & $.969$   & $<.001$ \\
Eating experience          & 1  & 3.55   & $.060$   & $<.001$ & 0.08  & $.779$   & $<.001$ \\
Gender                     & 1  & 0.91   & $.340$   & $<.001$ & 0.01  & $.911$   & $<.001$ \\
Group $\times$ Prompt      & 6  & 2.43   & $.024$   & .001    & 0.61  & $.723$   & $<.001$ \\
Group $\times$ Descriptor  & 22 & 14.43  & $< .001$ & .026    & 15.54 & $< .001$ & .028 \\
Prompt $\times$ Descriptor & 33 & 25.58  & $< .001$ & .067    & 28.41 & $< .001$ & .074 \\
Group $\times$ Prompt $\times$ Descriptor & 66 & 2.96 & $< .001$ & .016 & 3.18 & $< .001$ & .018 \\
\bottomrule
\end{tabular}
\end{table}

\subsection{Separating scale use from perceptual structure}

\begin{figure}[!ht]
\centering
\includegraphics[width=0.95\linewidth]{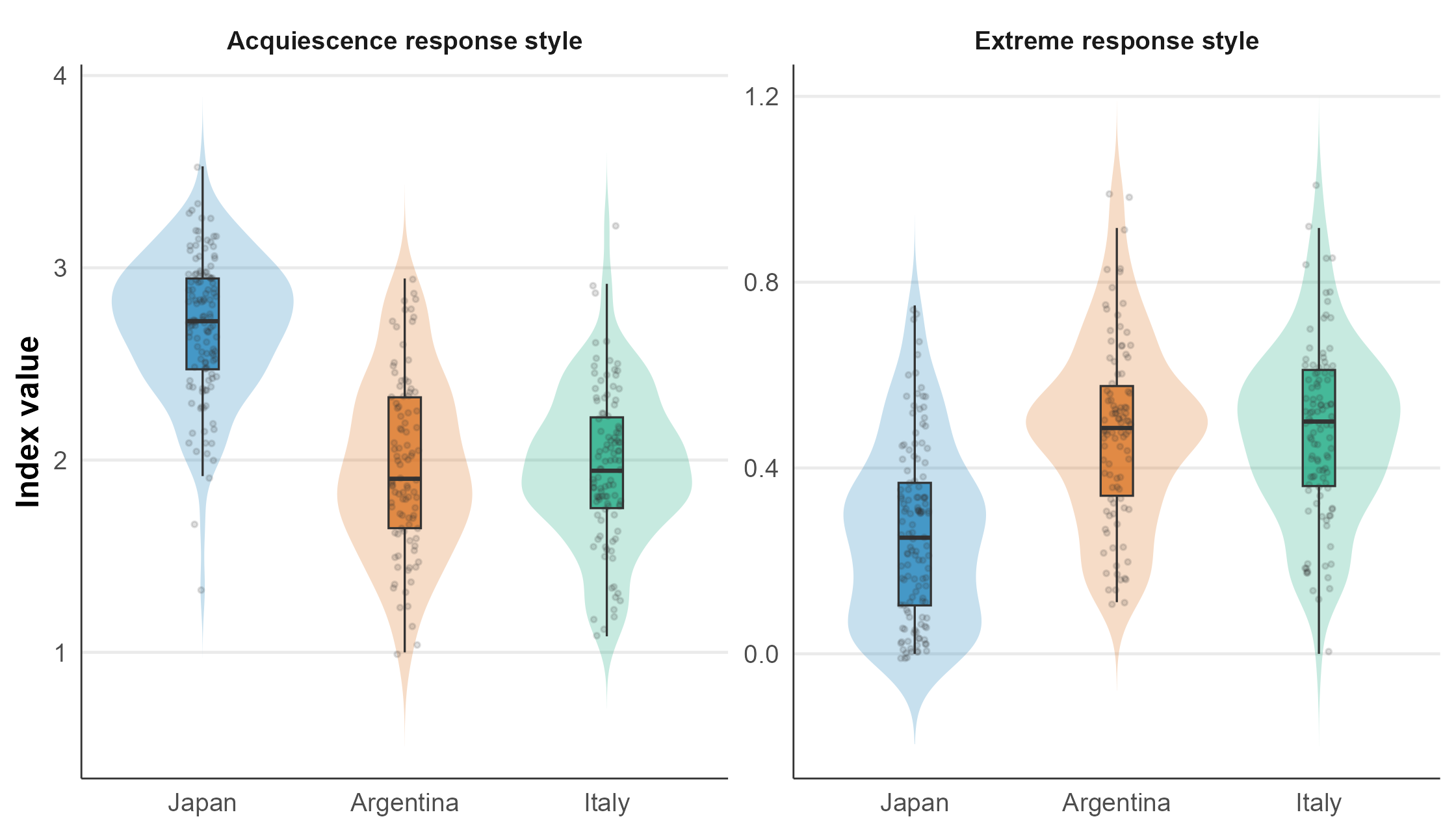}
\caption{Per-participant response-style distributions by cultural group. The left panel shows the
acquiescence response style index (participant mean across ratings), and the right panel shows the
extreme response style index (proportion of 1s and 5s); the two panels use independent y-axes because
the indices differ in unit. Japanese participants show the highest acquiescence but the lowest extreme
response, whereas the Argentine and Italian cohorts use lower but more extreme ratings.}
\label{fig:response-style}
\end{figure}

Response styles---habits of using a rating scale in a particular way regardless of item content---differ
markedly between the cohorts (Fig~\ref{fig:response-style}). Japanese participants have a substantially
higher acquiescence index, the mean of their 36 ratings (ARS $= 2.69$), than Argentine ($1.98$) or
Italian ($1.97$) participants, but a lower extreme-response index, the proportion of ratings at either
end of the scale (Japan $= 0.26$, Argentina $= 0.48$, Italy $= 0.47$); Kruskal--Wallis tests confirm both
differences (ARS: $p = 6.84 \times 10^{-35}$; ERS: $p = 8.61 \times 10^{-18}$). In short, Japanese
respondents used higher but more central ratings, the Argentine and Italian groups lower and more extreme
ones---exactly the profile that would inflate a group main effect without any difference in what was
heard.

To remove this component we re-expressed every rating relative to the participant's own mean and spread
(within-subject $z$-scoring), which places a participant who rates everything high on the same footing as
one who rates everything low. The right half of Table~\ref{tab:anova} shows the consequence, and it is
sharp. The main effect of group disappears completely ($F(2, 11{,}732) = 0.002$, $p = .998$, partial
$\eta^2 < .001$), and the \emph{group} $\times$ \emph{prompt} term disappears with it
($F(6, 11{,}732) = 0.61$, $p = .723$). By contrast, \emph{group} $\times$ \emph{descriptor} and the
three-way interaction remain strong, with effect sizes essentially unchanged from the raw model (partial
$\eta^2 = .028$ and $.018$). The corresponding z-scored mixed model
$(1 \mid \text{participant}) + (1 \mid \text{song})$ gives the same qualitative result (non-significant
\emph{group} $\times$ \emph{prompt}, $p = .714$; significant three-way interaction, $p < .001$). Most of
the raw country-level rating shift was therefore a scale-use difference; what remains is the
country-specific \emph{structure} of the mapping.

\begin{figure}[!ht]
\centering
\includegraphics[width=0.98\linewidth]{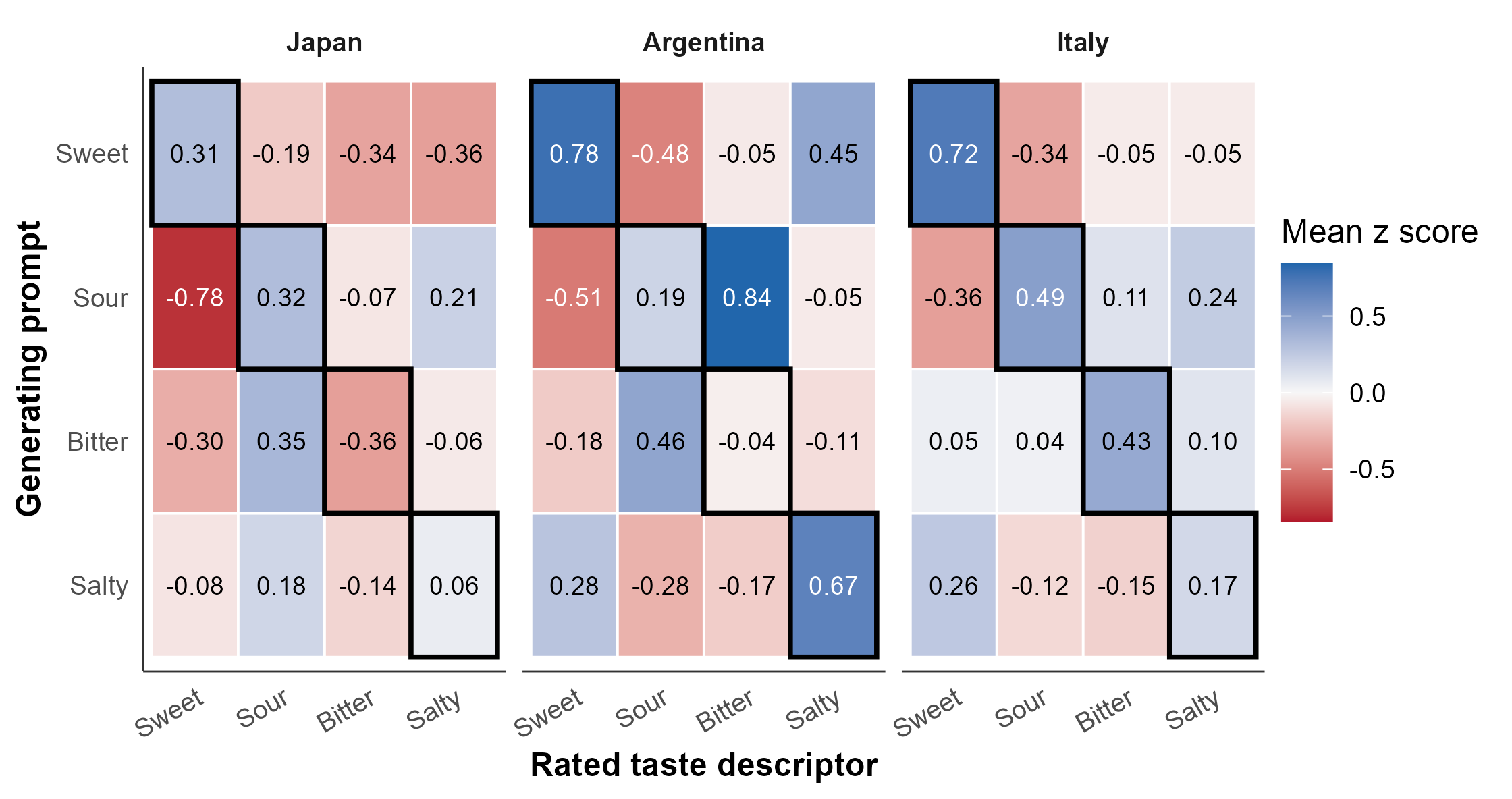}
\caption{Mean within-subject z-scored taste ratings in the Task~2 ANOVA sample. Rows are the prompt used
to generate the clip and columns the taste descriptor rated; black outlines mark the matching
prompt--descriptor cells, where a faithful correspondence would produce the highest values.}
\label{fig:taste-heatmap-z}
\end{figure}

That surviving structure is visible directly in the taste ratings (Fig~\ref{fig:taste-heatmap-z}). Each
panel is a prompt-by-descriptor grid, and a model that reproduced the intended correspondences would
place its highest values along the outlined diagonal. Italy comes closest, retaining a clear
sweet--sour--bitter diagonal with a weak salty cell. Argentina keeps sweet and salty but cross-maps sour
and bitter, the sour prompt pulling the bitter descriptor and the bitter prompt the sour one. Japan is
the most diffuse, with off-diagonal spillover around sour and no salty match. The raw ratings show the
same three arrangements superimposed on the country-level offset (Appendix~\ref{app:si}, Fig~A1),
which is what motivates working with the normalized values. The differences extend into the emotional and thermal
descriptors, again not identically. In the raw ratings, sweet clips are rated happier and warmer and
sour and bitter clips colder and more negative, especially in Argentina and Italy, while the Japanese
cohort shows a uniformly elevated affective profile across both positive and negative descriptors
(Appendix~\ref{app:si}, Fig~A2). Once that inflation is removed a structural
difference remains (Fig~A3): the sour prompt recruits negative-affect descriptors (anger, disgust, fear) most
strongly in Japan, the bitter prompt aligns chiefly with sadness in both Japan and Italy, and the sweet
prompt aligns with happy and hot in all three cohorts but to different degrees.

\subsection{Agreement between listeners}

The heatmaps average over listeners, so a cohort could show a clean mean structure while its individual
members disagreed completely. Before asking how the descriptor space is organized, we therefore ask how
much the listeners within a cohort agree about any single clip, using the mean pairwise distance (MPD,
the average absolute difference between two raters of the same cell, $0$ indicating perfect agreement)
and Krippendorff's $\alpha$ ($1$ perfect agreement, $0$ chance).

Cell-level disagreement is moderate and comparable across cohorts (mean MPD $0.27$--$0.29$ on a $0$--$1$
scale). The aggregated $\alpha$ values are all well below the conventional $0.667$ threshold, as expected
for single-trial perceptual ratings of AI-generated audio by untrained listeners, but differ meaningfully
across cohorts: Argentina ($\alpha = 0.192$) and Japan ($\alpha = 0.172$) reach roughly twice the
agreement of Italy ($\alpha = 0.103$); an ordinal-metric variant falls within $\pm 0.01$ of these
(cohort table in Appendix~\ref{app:si}, Table~A16; the per-cell breakdown is
archived with the data). Italy is
the noisiest
mainly because its bitter cell collapses to chance: Italian listeners rated the bitter prompt almost
identically on all four taste descriptors (Appendix~\ref{app:si}, Fig~A1), leaving little systematic signal for raters
to agree on. Agreement is thus low everywhere and lowest in the cohort with the
cleanest average diagonal, so the cross-cultural differences reported here are group-level regularities
that emerge from noisy individual judgments rather than consensus perceptions.

\subsection{Factor structure across cultures}

If the three cohorts organize the descriptor space differently, that difference should appear not only in
the mean ratings but in how the twelve descriptors covary. Exploratory factor analysis groups descriptors
that move together into a small number of latent dimensions, and Tucker's congruence coefficient measures
how similar two such solutions are ($1.0$ identical; above $0.95$ conventionally treated as equivalent,
$[0.85, 0.95)$ as fair similarity).

Parallel analysis returns four factors for Japan, three for Argentina, and four for Italy. When the
solutions are truncated to three factors for comparison, congruence remains low: $0.777$ for
Japan--Italy, $0.613$ for Japan--Argentina, and $0.460$ for Argentina--Italy
(Table~\ref{tab:congruence}), all below the equivalence threshold and two of the three below the lower
bound of fair similarity. A participant bootstrap returns wide intervals whose upper bounds still fall
short of $0.95$, so the low congruence is a stable feature of the data rather than a small-sample
artifact, consistent with the exploratory status of the comparison.

\begin{figure}[!ht]
\centering
\includegraphics[width=0.98\linewidth]{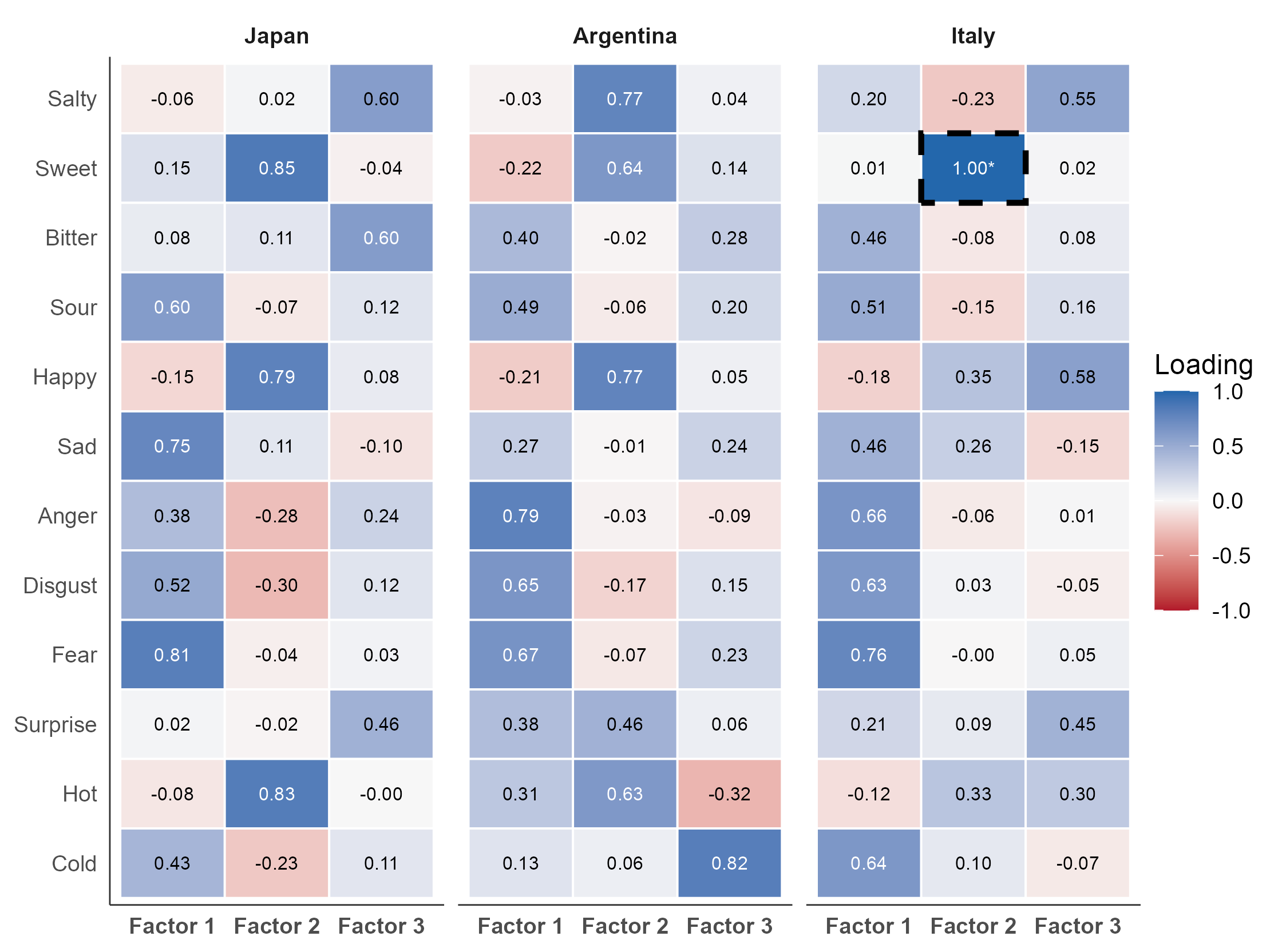}
\caption{Three-factor loading patterns by cohort. Blue indicates positive loadings and red negative.
The dashed-outline cell in the Italian solution marks a Heywood case (loading $\approx 1.00$, residual
variance close to zero), marking the Italian three-factor solution as exploratory.}
\label{fig:factor-loadings}
\end{figure}

Fig~\ref{fig:factor-loadings} shows why the congruence values are so low. In Japan, sweet, happy, and hot
load together on a positive factor, while negative affect clusters with sour, and salty groups with bitter and
surprise. In Argentina, salty, sweet, and happy collapse into a warm--pleasant factor, while anger,
disgust, fear, sour, and bitter load together on a negative-affect factor, with cold defining a separate
third dimension, and hot and surprise loading on both of the first two factors rather than on either
alone. In Italy, sweet is nearly a single-item factor, while salty, happy, and surprise form a separate
dimension and negative affect loads with sour, bitter, and cold. These are different semantic
organizations of the same descriptor set, not small rotations of a common structure: in Argentina the
organizing principle is affective valence, in Italy it is closer to the tastes themselves.

\begin{table}[!ht]
\centering
\caption{Mean diagonal Tucker congruence coefficients between the three cohort-specific factor solutions
(three-factor truncation, oblimin rotation, maximum-likelihood extraction). Values above $0.95$ are
usually interpreted as factor equivalence and values in $[0.85, 0.95)$ as fair similarity; every
off-diagonal value here falls below both thresholds.}
\label{tab:congruence}
\small
\begin{tabular}{lrrr}
\toprule
          & Japan & Argentina & Italy \\
\midrule
Japan     & 1.000 & 0.613 & 0.777 \\
Argentina & 0.613 & 1.000 & 0.460 \\
Italy     & 0.777 & 0.460 & 1.000 \\
\bottomrule
\end{tabular}
\end{table}

\subsection{Acoustic characterization of the stimuli}
\label{sec:acoustic}

Everything so far has treated the clips as fixed and the listeners as the variable of interest. This
last section reverses the perspective and asks what the clips actually sound like, since a cross-cultural
difference is only interpretable once we know what acoustic evidence the three cohorts were working from.
We characterized the 100 fine-tuned clips (25 per prompt) on eight acoustic descriptors and a
Plomp--Levelt sensory-dissonance index (Section~\ref{sec:acoustic-method}), read as relative patterns across prompts.

\begin{figure}[!ht]
\centering
\includegraphics[width=0.98\linewidth]{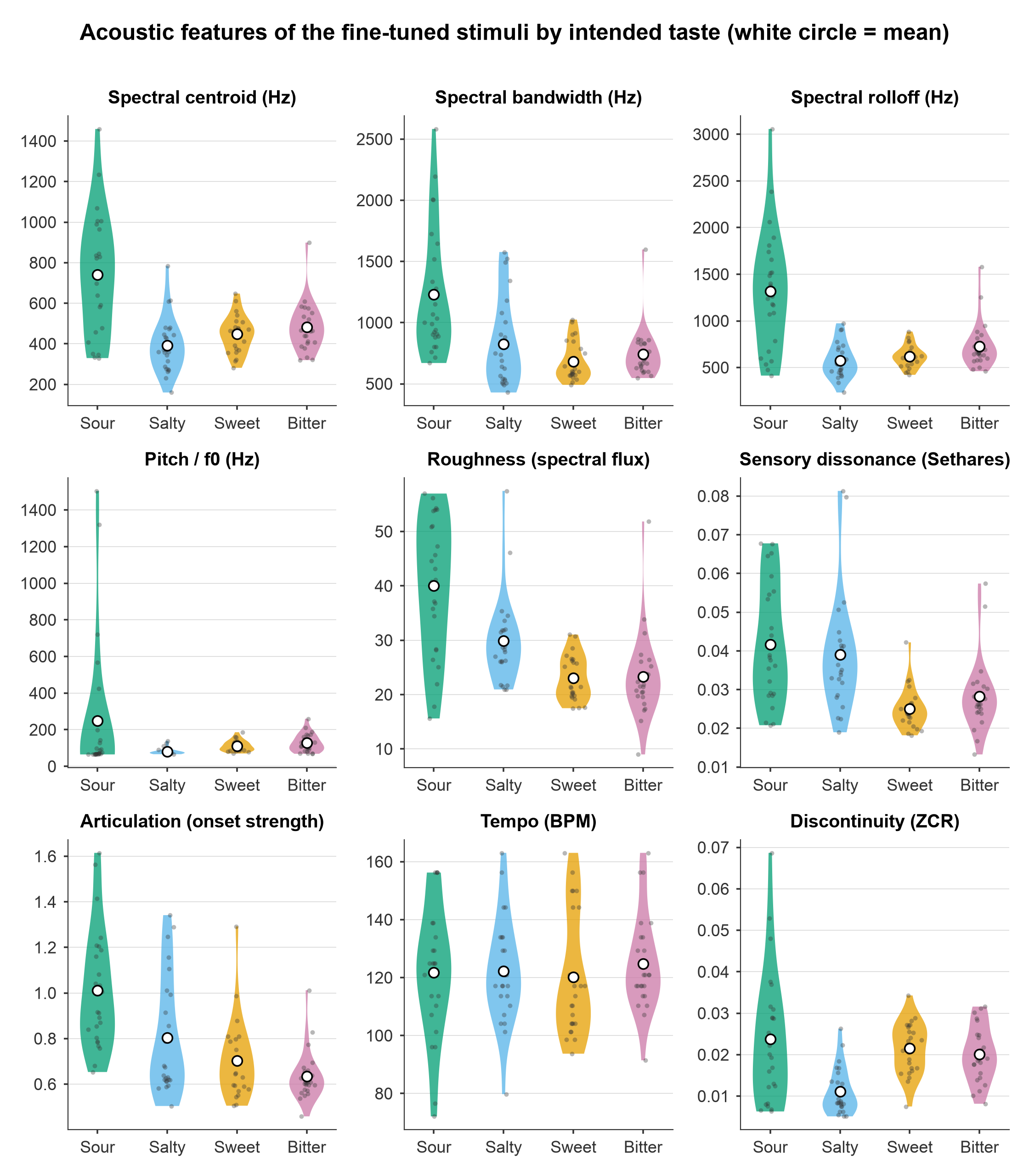}
\caption{Acoustic descriptors of the 100 fine-tuned clips by intended taste prompt (white circle marks
the mean). Sour is the lone acoustic outlier; sweet, salty, and bitter occupy a common dark, smooth,
low-pitched region. Tempo and zero-crossing rate are essentially non-discriminating across prompts.}
\label{fig:acoustic}
\end{figure}

Two results stand out (Fig~\ref{fig:acoustic}). First, the fine-tuned model gave a distinctive acoustic
signature to only one taste. Sour clips are markedly brighter, sharper, higher-pitched, rougher, and
more articulated than the others (centroid $741$ versus $393$--$482$ Hz; pitch $250$ versus $81$--$127$
Hz; spectral flux $40$ versus $23$--$30$), whereas sweet, salty, and bitter collapse into a single dark,
smooth, low-pitched region: in within-feature $z$-scored descriptor space those three lie only
$0.4$--$1.1$ apart, while sour sits $2.9$--$3.2$ from every one of them. Salty in particular has the
lowest centroid and pitch of any prompt and none of the crisp, percussive profile of designed salty
sounds \cite{deng_composing_palate_2026}. This is a concrete stimulus-level reason for the weak salty
match seen in both tasks, and it complements the intrinsic-ambiguity account developed in the Discussion.

Second, although the model did not actually make its \emph{bitter}-prompted clips dissonant (they are
among the most consonant of the four prompts), sensory dissonance is nonetheless a shared route to
\emph{both} sour and bitter perception. Across the 100 clips, the dissonance index correlates positively
with pooled perceived sour ($r = .25$, $p = .012$) and bitter ($r = .30$, $p = .003$) and negatively with
sweet ($r = -.47$, $p < .001$); spectral roughness shows the same pattern more strongly (sour $r = .31$,
bitter $r = .41$, sweet $r = -.57$, all $p \le .002$). Dissonance is thus a single perceptual axis that
does not separate the two negative tastes---precisely the substrate that a sour--bitter cross-map
requires.

The prompt that generated a clip is the designer's intention, not necessarily what listeners heard.
Relabelling each clip by its perceived dominant taste (the highest of its four response-style-corrected
taste ratings) redistributes the prompts substantially: only $39$ of the $100$ clips were heard as the
taste they were generated for (Table~\ref{tab:confusion}). Sweet is the only prompt reliably recovered
($18/25$); the bitter prompt almost never is ($2/25$), its clips heard instead as sour ($11$) or
sweet ($10$). Grouped by perceived rather than intended taste, the acoustic contrasts become cleaner
(Appendix~\ref{app:si}, Fig~A7): clips heard as sweet are the softest and most
consonant (roughness
and dissonance
about $0.6$ SD below the grand mean), those heard as salty the most articulated and widest-band, and
those heard as sour or bitter the roughest and most dissonant, with perceived bitter the most extreme. A
$k$-means analysis of the perceived-taste ratings finds no clean four-way structure (best silhouette at
$k = 2$, $0.33$ versus $0.27$ at $k = 4$), so perceived bitter and sour share a single rough, dissonant
region rather than forming separate perceptual types.

Because each clip is rated by only about ten participants ($5$ to $26$ across clips), this
dominant-taste assignment is noisy: bootstrapping the raters reproduces the assigned taste on a median of
$79\%$ of resamples, but for nine of the $100$ clips the dominant taste flips in more than half. We
therefore treat the clip-level relabeling as exploratory rather than a stable per-clip decoding of taste.

\begin{table}[!ht]
\centering
\caption{Intended prompt (rows) versus perceived dominant taste (columns) for the $100$ fine-tuned clips.
Perceived taste is the highest of each clip's four response-style-corrected taste ratings. Only the $39$
on-diagonal clips (bold) were heard as the taste they were generated for; the bitter prompt is heard as sour
or sweet far more often than as bitter.}
\label{tab:confusion}
\small
\begin{tabular}{lrrrr}
\toprule
Prompt $\downarrow$ / Heard $\rightarrow$ & Sweet & Sour & Salty & Bitter \\
\midrule
Sweet  & \textbf{18} & 2  & 3 & 2 \\
Sour   & 1  & \textbf{10} & 8 & 6 \\
Salty  & 6  & 6  & \textbf{9} & 4 \\
Bitter & 10 & 11 & 2 & \textbf{2} \\
\bottomrule
\end{tabular}
\end{table}

\begin{figure}[!ht]
\centering
\includegraphics[width=\linewidth]{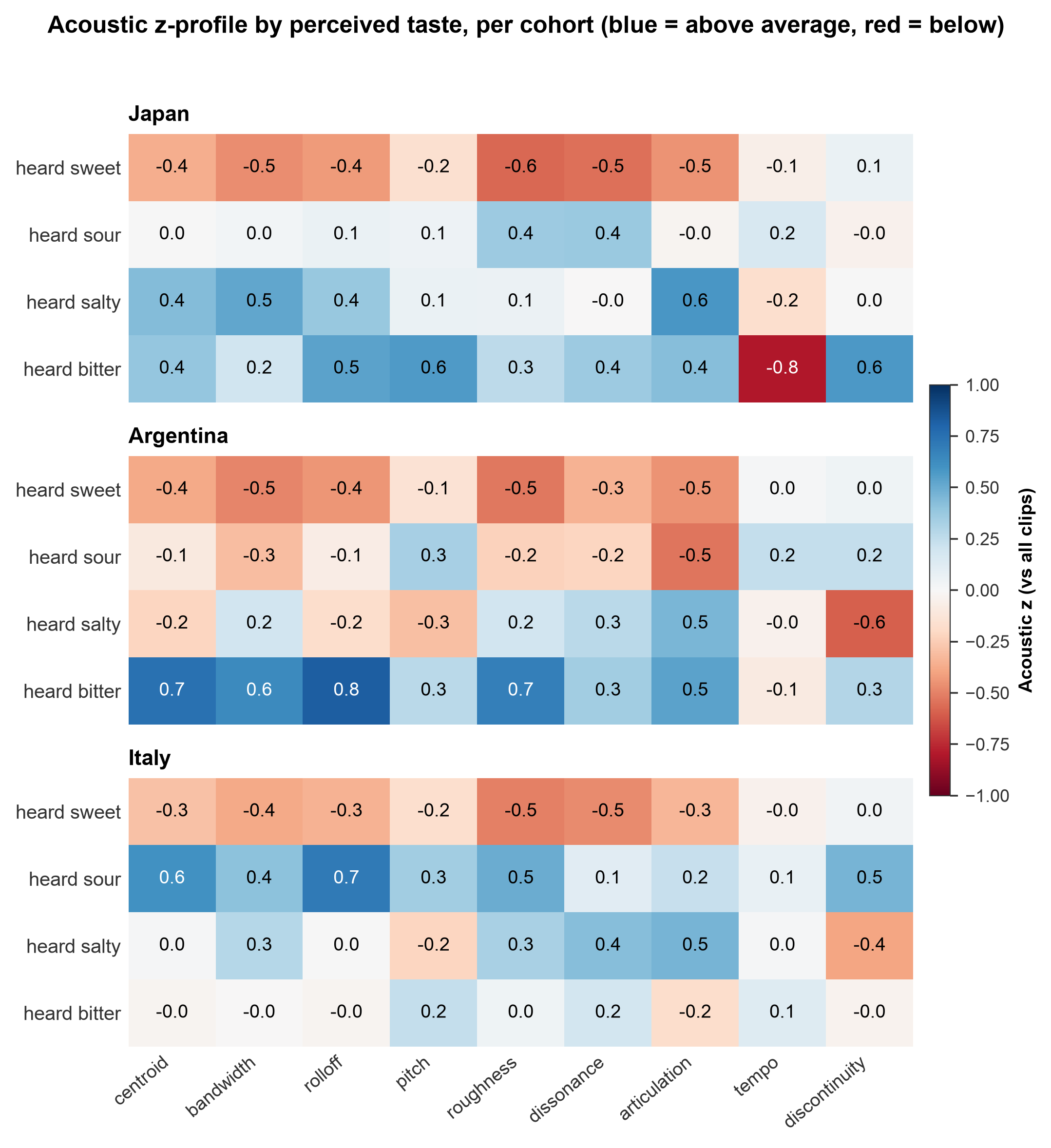}
\caption{Acoustic $z$-profile (relative to all $100$ clips) of each \emph{perceived} taste, derived
separately within each cohort. Clips heard as sweet are soft and consonant in every cohort; the rough,
bright, dissonant region is heard as sour in Italy and Japan but as bitter in Argentina, and Japanese
listeners rarely assign bitter at all. Blue marks features above the cross-clip average, red below.}
\label{fig:cohort-acoustic}
\end{figure}

Deriving perceived taste separately within each cohort brings the listener-side and stimulus-side
analyses together, and shows that the perceptual mapping is shared for sweetness but reorganized for the
negative tastes (Fig~\ref{fig:cohort-acoustic}). Clips heard as sweet have the same soft, consonant
profile everywhere (cross-cohort $r = 0.93$--$0.99$; roughness predicts reduced perceived sweetness in
every cohort, $r = -0.42$ to $-0.51$, all $p < .05$). What differs is the taste descriptor attached to the
rough, bright, dissonant clips: within-cohort correlations between clip-level spectral roughness and the
perceived taste ratings tie that region to sourness in Italy (roughness--sour $r = .36$, $p < .05$;
roughness--bitter $r = .07$), to bitterness in Argentina ($r = .07$ and $r = .40$, $p < .05$), and to both
in Japan ($r = .30$ and $r = .29$, both $p < .05$; full table in Appendix~\ref{app:si}, Table~A20). The
Argentine
bitter clips are the brightest and roughest of all ($\approx 0.7$ SD above the mean), the very profile
Italian and Japanese listeners assign to sourness, and Japanese listeners almost never choose bitter as
the dominant descriptor ($2$ of $99$ clips, against $44$ heard as sour). The per-cohort acoustic profiles
of perceived sweet are thus nearly identical, whereas those of perceived sour and bitter are uncorrelated
or negatively correlated across cohorts. The sour--bitter swap and the diffuse, sour-centred Japanese map
seen in the rating structure are therefore confirmed in the stimulus acoustics: the three cohorts were
listening to the same sounds and attaching different taste descriptors to them.

\FloatBarrier
\section{Discussion}

The present study examined whether AI-mediated auditory--gustatory correspondences generalize across cultures and, if not, what mechanisms account for the observed differences. The results indicate that they do not generalize uniformly. Cross-cultural variation emerged less as a simple shift in overall ratings than as differences in how listeners organized the relationship between musical stimuli and taste descriptors. Two complementary processes appear to underlie this pattern. First, culturally specific response styles influenced the absolute level of semantic ratings. Second, even after accounting for these response-style differences, the semantic organization of auditory--gustatory correspondences remained different across countries.

At the preference level, the findings provide a partial replication of the original Italian study. Participants from Italy and Argentina consistently preferred music generated by the fine-tuned MusicGen model over the original model, whereas Japanese participants did not. This pattern suggests that the benefits of fine-tuning are not equally transferable across listening cultures. By contrast, the salty prompt performed poorly in all three cohorts, indicating a limitation of the generated stimuli rather than a purely cultural effect.

The cross-cultural comparison also refines the conclusions of the original Italian study. Spanio et al.~\cite{spanio_frontiers_2025} demonstrated that a fine-tuned generative model could produce musical stimuli perceived as more coherent with taste prompts than those generated by the original model. The present findings show that this conclusion depends partly on cultural context. Although the overall advantage of the fine-tuned model remains evident in two of the three countries, the comparison reveals substantial heterogeneity that is obscured when responses are pooled across listeners. Similarly, the clear prompt--descriptor structure observed in the original Italian sample does not remain invariant across cultures but reorganizes in different ways across the three cohorts.

A second contribution concerns measurement. Raw semantic ratings alone would suggest large differences between countries, largely because Japanese participants systematically used higher but less extreme ratings than the Argentine and Italian cohorts. Once these response-style differences were taken into account through within-subject normalization, the overall country effect largely disappeared. However, the interactions describing the relationship between prompts and semantic descriptors remained essentially unchanged. This distinction is important because it shows that cross-cultural variation cannot be reduced to differences in scale use. Instead, listeners from different cultures appear to organize the semantic relationships between music and taste differently, even after response-style effects are removed.

Taken together, these findings suggest that cross-cultural variation in AI-mediated sonic seasoning operates at two distinct levels. One concerns how participants use semantic rating scales, whereas the other concerns how auditory information is mapped onto gustatory concepts. Distinguishing these two sources of variation is essential for interpreting cross-cultural evaluations of generative music systems, since apparent differences between populations may reflect changes in semantic organization rather than simple differences in response behaviour.

The cross-cultural reorganization observed in the present study also suggests that the relationship between sound and taste is mediated by multiple cognitive processes rather than by direct perceptual associations alone. Deng et al.~\cite{deng_composing_palate_2026} identified six recurrent strategies that listeners use when assigning taste meanings to sound: reliance on sensory features, subjective preference, personal experiential analogies, affective responses, somatic reactions, and cognitive abstractions such as metaphor or internal scaling. Among these, only the direct use of sensory features is likely to be relatively independent of cultural experience. The remaining strategies are inherently shaped by individual experience and, consequently, by the musical, culinary, and linguistic environment in which listeners are embedded.

Our findings are consistent with this interpretation. The differences observed across countries do not necessarily reflect fundamentally different perceptual mechanisms; rather, they suggest that listeners rely on different combinations of these mediating strategies when interpreting the same musical stimuli. Preferences, culturally familiar foods, emotional associations, and linguistic categories all contribute to how a sound is ultimately translated into a taste concept. From this perspective, the cross-cultural reorganization identified in the present study represents the population-level expression of cognitive processes that operate at the level of individual listeners.

Although Deng et al.~\cite{deng_composing_palate_2026} discussed this possibility, their conclusions were necessarily limited to a single participant population. By extending the same general framework to three culturally distinct cohorts, the present study provides direct evidence that these mediating processes are themselves influenced by cultural context. The following sections examine this interpretation from three complementary perspectives: the perceptual substrate shared across cultures, the contribution of musical enculturation, and the specific characteristics of the Argentine cohort, for which directly comparable taste--music research is available.

\subsection{A shared perceptual substrate}

The cross-cultural differences observed in this study rest on a perceptual organization that is nevertheless shared across the three cohorts. The clearest example concerns the relationship between sourness and bitterness. Although the two tastes were not labelled consistently across countries, listeners in all three cohorts associated them with the same region of the acoustic space, characterized by greater roughness and sensory dissonance. The cross-cultural variation therefore concerns how these acoustic cues are interpreted rather than whether or not they are perceived.

This interpretation is consistent with previous work on taste--music correspondences. Musical improvisations prompted by taste descriptors consistently associated both sourness and bitterness with dissonant musical structures, while distinguishing them through secondary features such as pitch and articulation \cite{mesz_taste_music_2011}. Likewise, studies using isolated scales and chords have shown that musical dissonance provides one of the principal dimensions organizing taste judgments, supporting the view that sourness and bitterness occupy neighbouring regions within a common perceptual space rather than being defined by entirely distinct acoustic cues \cite{taitz_taste_scales_2021}. Our acoustic analysis extends these observations to AI-generated music. Across the generated stimuli, sensory dissonance and spectral roughness predicted both perceived sourness and perceived bitterness, while showing the opposite relationship with sweetness. Dissonance therefore appears to provide a robust cue for distinguishing pleasant from unpleasant taste qualities, but not for separating the two negative tastes from one another.

The ambiguity lies in listeners' perception rather than in the prompt used to generate the stimuli. The fine-tuned MusicGen model did not consistently generate more dissonant music for bitter prompts than for the other tastes. Instead, whenever roughness and dissonance were present, listeners interpreted those cues as either sourness or bitterness depending on the semantic organization characteristic of their cultural group. A similar discrepancy between intended and perceived taste was reported by Deng et al.~\cite{deng_composing_palate_2026}, suggesting that this mismatch reflects a broader property of AI-generated sonic seasoning rather than a limitation specific to the present study.

The same conclusion emerges when the clips are classified according to the taste most frequently perceived by listeners rather than according to the prompt that generated them. Under this representation, clips perceived as sweet consistently occupied the smoothest and most consonant region of the acoustic space, whereas clips perceived as sour or bitter clustered within a common region characterized by greater roughness and dissonance. Rather than contradicting previous studies, these results show that AI-generated stimuli reproduce the same broad auditory--gustatory organization previously observed with manually designed musical material.

Saltiness followed a different pattern. It was the weakest and least consistently identified taste in every cohort, suggesting that this limitation is unlikely to arise solely from cultural interpretation. Previous work has shown that the auditory representation of saltiness remains comparatively elusive and is not associated with a single dominant acoustic profile \cite{wang_saltiness_2021,deng_composing_palate_2026}. Our acoustic analysis points to the same conclusion. Compared with the other prompts, the generated salty excerpts lacked the bright, crisp, and articulated acoustic features that have previously been associated with perceived saltiness, leaving listeners in all three countries with relatively little acoustic evidence on which to base a consistent salty judgment.

Overall, these findings suggest that cultural differences emerge from the interpretation of shared perceptual cues rather than from fundamentally different auditory representations. The acoustic information conveyed by the generated music remains broadly similar across cultures, whereas the semantic labels attached to that information are reorganized according to culturally acquired patterns of interpretation. The next section considers how musical enculturation may contribute to this reorganization.

\subsection{A musical-culture perspective}


The shared perceptual substrate identified above does not imply that listeners attach the same meaning to identical musical cues. Instead, the present findings suggest that cultural variation emerges when those cues are interpreted and translated into semantic categories. Musical enculturation provides a natural explanation for this process.
Listeners acquire expectations about harmony, timbre, rhythm, and expressive meaning through prolonged exposure to the musical traditions of their own culture. These expectations influence both emotional responses to music and the semantic categories listeners use to describe what they hear \cite{balkwill_japanese_2004,klarlund_worlds_apart_2023,li_cross_cultural_biases_2025,di_stefano_prokofiev_2024}. Consequently, the same acoustic features may evoke similar low-level perceptual responses while being interpreted differently across cultures.
This interpretation fits the pattern observed here. The Italian cohort retained the clearest correspondence with the intended taste structure, whereas the Japanese and Argentine cohorts reorganized the same acoustic information in different ways. Rather than reflecting different perceptual mechanisms, these differences are more plausibly understood as culturally specific interpretations of a common perceptual substrate.

\subsection{Interpreting the Argentine taste--music map}\label{sec:interpret-argentine}

The Argentine cohort deserves particular attention because previous studies have examined taste--music correspondences within the same cultural context, providing an opportunity to compare the present findings with an existing behavioural literature. The most distinctive result was the systematic sour--bitter cross-map. Unlike the Italian and Japanese cohorts, the bright, rough, and dissonant musical excerpts that were predominantly perceived as \emph{sour} in those countries were more often judged as \emph{bitter} by Argentine listeners.

This pattern is not entirely consistent with previous behavioural work. Musical improvisations prompted by taste descriptors showed that Argentine musicians reliably distinguished sourness from bitterness, assigning different musical characteristics and emotional associations to the two tastes \cite{mesz_taste_music_2011}. The present findings therefore do not suggest that the underlying perceptual substrate differs in Argentina. Rather, they indicate that the same acoustic information is interpreted differently when listeners assign taste descriptors to AI-generated music.

One possible explanation concerns the acoustic properties of the generated stimuli themselves. As discussed in Section~\ref{sec:acoustic}, the fine-tuned MusicGen model did not produce a clear acoustic separation between the sour and bitter prompts. If both prompts occupy a common rough and dissonant region of the acoustic space, listeners may first identify a generic unpleasant auditory quality before selecting the taste descriptor that best matches their own semantic representation. The Argentine sour--bitter cross-map may therefore reflect an interaction between stimulus ambiguity and culturally specific semantic organization rather than either factor in isolation.

The factor analysis reinforces this interpretation. Unlike the Italian and Japanese solutions, the Argentine factor structure was organized primarily around affective dimensions rather than individual tastes. Sourness and bitterness loaded together with anger, disgust, and fear, whereas sweetness and saltiness clustered with happiness (Fig.~\ref{fig:factor-loadings}). A similar organization has been reported previously in studies conducted with Argentine participants. In musical improvisations, bitter music was associated with pain and sadness, whereas sour music evoked unpleasantness, fear, and cruelty, while saltiness was linked to joyful emotional content \cite{mesz_taste_music_2011}. Likewise, Galmarini et al.~\cite{galmarini_impact_2021} showed that background music influenced the emotional experience of coffee more strongly than its perceived taste, suggesting that emotional transfer may play a particularly prominent role in this cultural context.

Language may contribute to the same pattern. Taste descriptors are not emotionally neutral labels but concepts closely linked to affective meaning \cite{avery_taste_metaphors_2022}. Moreover, the semantic associations evoked by these concepts vary across languages and cultures rather than following a single universal organization \cite{wan_crosscultural_2014}. Because participants completed the questionnaire in their native language, part of the observed reorganization may reflect differences in the semantic representation of taste concepts themselves. This interpretation is consistent with the \emph{cognitive abstraction} strategy described by Deng et al.~\cite{deng_composing_palate_2026}, whereby listeners rely not only on acoustic information but also on culturally acquired semantic knowledge when assigning taste descriptors to sound.

The dual loading of the descriptors \emph{hot} and \emph{surprise} provides an additional indication that affective meaning contributes to the Argentine semantic structure. Both descriptors loaded on more than one factor, suggesting that they cannot be interpreted as belonging to a single perceptual dimension. One possible explanation is that these concepts evoke multiple culturally meaningful associations. For example, \emph{hot} may simultaneously reflect the contemplative warmth associated with the \emph{mate amargo} tradition and the convivial atmosphere surrounding the \emph{asado}. Likewise, the dual loading of \emph{surprise} is compatible with the bittersweet emotional character often associated with tango and Argentine folklore, where positive and negative emotions frequently coexist rather than opposing one another \cite{mesz_taste_music_2011}. These interpretations remain speculative, but they provide culturally grounded hypotheses that are consistent with the semantic organization observed in the present data.

\subsection{Implications and limitations}

The present findings have implications for both the evaluation and development of generative music systems. A model that performs well within one cultural context should not be assumed to generalize simply because its prompts are linguistically interpretable or its outputs appear perceptually coherent to a single population. Cross-cultural validation should distinguish genuine differences in perceptual organization from differences in response style, and should account for the repeated-measures structure of semantic rating data. More broadly, the results reinforce the need for culturally diverse training and evaluation datasets in generative music models, rather than assuming that a single cultural benchmark provides adequate validation \cite{kanatas_culturemert_2025}.

Several limitations should nevertheless be acknowledged. First, although the three cohorts were selected to maximize cultural diversity, they differed in demographic characteristics, and not all between-country differences can therefore be interpreted as purely cultural. Second, the response-style correction adopted here was intentionally simple. More comprehensive approaches based on measurement invariance or latent-variable modelling would provide a stronger assessment of cross-cultural comparability. Third, the factor analysis was exploratory, and the mixed-effects models included random intercepts but not random slopes. The resulting factor structures should therefore be interpreted as evidence of cross-cultural organization rather than definitive measurement models.

Fourth, the stimulus-generation pipeline was anchored in English: MusicGen was conditioned on English taste prompts through its English-trained T5 text encoder. Because the identical clips were presented to all three cohorts, this generation-prompt language was constant across countries and cannot, on its own, explain why the fine-tuned model was preferred in Italy and Argentina but not in Japan. The bias is also narrow. It is architectural, limited to the text-conditioning interface: the model's language representation is skewed toward English, whereas its learned audio representation (the space of music it can encode and generate) shows no evident incompleteness or comparable bias, even though its training distribution is weighted toward Western music \cite{kanatas_culturemert_2025}. This leaves only the question of how each English taste prompt is mapped onto a conditioning vector. Studies of large language models show that prompt language can shift a model's outputs and activate language-specific, often culturally specific, knowledge \cite{wang2025multilingual}, and that the size and direction of the effect depend on the language involved \cite{mondshine2025beyond,yin2024respect}. Because these studies concern question answering, inference, and text generation rather than text-to-music synthesis, the parallel holds only by analogy. Subject to that caveat, English prompting may have nudged the taste-to-music mapping toward a Western construal of the four taste prompts. Any such effect is a shared property of the stimuli, not a cohort-specific measurement artifact, and would produce a difference between cohorts only through its interaction with listeners' culturally acquired semantic organization. We advance this as a hypothesis, not a demonstrated cause. It is distinct from, though complementary to, the effect of the native-language questionnaire on participants' own responses, which we discuss (with the closer linguistic proximity of Italian and Spanish to the taste concepts) in Section~\ref{sec:interpret-argentine}.

Finally, the sour--bitter reorganization observed in the Argentine cohort should not be interpreted as evidence of a purely cultural effect independent of the stimuli themselves. The acoustic distinction between the sour and bitter prompts generated by the fine-tuned model was limited, and this ambiguity almost certainly contributed to the observed cross-map. At the same time, the cohort-specific perceived-taste analysis (Section~\ref{sec:acoustic}) showed that the same bright, rough, and dissonant clips were consistently interpreted as \emph{sour} in Italy and Japan but as \emph{bitter} in Argentina. Holding the acoustic material essentially constant while the semantic label changes strongly suggests that the observed reorganization reflects an interaction between stimulus ambiguity and culturally specific interpretation rather than either mechanism alone.

\section{Conclusion}

This study provides the first cross-cultural evaluation of AI-mediated sonic seasoning across Argentina, Italy, and Japan. Although the fine-tuned MusicGen model generalized successfully to two of the three cohorts, its performance was not culturally invariant. More importantly, the cross-cultural differences observed here extended beyond overall model preference. Once response-style effects were taken into account, listeners from different countries continued to organize the semantic relationships between music and taste in distinct ways.

These findings suggest that auditory--gustatory correspondences combine shared perceptual regularities with culturally specific semantic interpretation. AI-generated music therefore appears capable of reproducing broad crossmodal associations, but not necessarily the way those associations are organized within different cultural contexts.

From a methodological perspective, the results also show that cross-cultural evaluations of generative models should distinguish response styles from perceptual organization. Without this distinction, differences in semantic ratings may overestimate cultural divergence or obscure genuinely shared perceptual structure.

More broadly, the present study illustrates how generative AI can serve not only as a creative technology but also as a tool for investigating multisensory perception. Evaluating AI-generated stimuli across culturally diverse populations offers a promising approach for understanding both the generality and the cultural specificity of human crossmodal correspondences.

\section*{Acknowledgments}
This work was partially funded by the European Union -- NextGenerationEU, under the
National Recovery and Resilience Plan (PNRR).

\section*{Data and code availability}
The survey data, the audio stimuli and the complete analysis code are openly available.
Code: \url{https://github.com/matteospanio/multimodal-survey-analysis}. Data and stimuli:
\href{https://doi.org/10.5281/zenodo.20841611}{10.5281/zenodo.20841611} (CC BY 4.0).

\clearpage
\bibliographystyle{unsrt}
\bibliography{references}

\clearpage
\appendix
\setcounter{table}{0}
\setcounter{figure}{0}
\renewcommand{\thetable}{A\arabic{table}}
\renewcommand{\thefigure}{A\arabic{figure}}

\section{Supplementary methods, tables and figures}
\label{app:si}

This appendix collects the supporting analyses referenced in the main text, in the same order as
the Results. Each table and figure is followed by a short \emph{How to read it} note stating what
the numbers mean, so a subsection can be consulted on its own. Every value is produced by the
analysis pipeline archived with the paper, from the same data and the same model specifications as
the main text.

\subsection{Sample composition and group equivalence}

The three cohorts were recruited independently, so before any cross-cultural comparison we
describe them in full and test whether they are comparable on background variables. They are
not: age, gender, musical experience and food experience all differ, which is why the three
experience and gender variables enter every rating model in the manuscript as covariates.

\begin{table}[H]
\centering
\caption{Extended demographic breakdown by cohort. Categorical rows give the count with the within-cohort percentage in parentheses.}
\small

\begin{tabular}[t]{llrrr}
\toprule
  & Level & Japan & Argentina & Italy\\
\midrule
Participants & $N$ & 140 & 104 & 117\\
Age (years) & mean $\pm$ SD & 27.2 $\pm$ 13.3 & 42.4 $\pm$ 15.0 & 34.2 $\pm$ 14.7\\
 & median (range) & 21 (18--77) & 42 (18--93) & 28 (19--75)\\
Completion time (min) & median & 8.0 & 8.0 & 8.0\\
Gender & Male & 40 (28.6) & 51 (49.0) & 63 (53.8)\\
 & Female & 96 (68.6) & 52 (50.0) & 51 (43.6)\\
 & Other & 1 (0.7) & 0 (0.0) & 2 (1.7)\\
 & Not specified & 3 (2.1) & 1 (1.0) & 1 (0.9)\\
Musical experience & Professional & 3 (2.1) & 26 (25.0) & 38 (32.5)\\
 & Amateur & 43 (30.7) & 45 (43.3) & 41 (35.0)\\
 & Not-experienced & 94 (67.1) & 33 (31.7) & 38 (32.5)\\
Food experience & Professional & 8 (5.7) & 5 (4.8) & 9 (7.7)\\
 & Amateur & 27 (19.3) & 42 (40.4) & 46 (39.3)\\
 & Not-experienced & 105 (75.0) & 57 (54.8) & 62 (53.0)\\
Listening device & Headphones & 83 (59.3) & 73 (70.2) & 77 (65.8)\\
 & Speakers & 47 (33.6) & 25 (24.0) & 34 (29.1)\\
 & HiFi system & 1 (0.7) & 4 (3.8) & 4 (3.4)\\
 & Other & 9 (6.4) & 2 (1.9) & 2 (1.7)\\
Hearing impairment & Yes & 5 (3.6) & 4 (3.8) & 6 (5.1)\\
 & No & 135 (96.4) & 100 (96.2) & 111 (94.9)\\
Taste impairment & Yes & 2 (1.4) & 3 (2.9) & 3 (2.6)\\
 & No & 138 (98.6) & 101 (97.1) & 114 (97.4)\\
\bottomrule
\end{tabular}
\end{table}
\noindent\emph{How to read it.} Read down a column for the profile of one cohort, across a row to compare cohorts on one characteristic. The Japanese cohort is the youngest and the most female; the Argentine cohort the oldest and the most gender-balanced.

\begin{table}[H]
\centering
\caption{Group equivalence tests on background variables. Fisher's exact tests use Monte Carlo simulation ($B = 10^5$) because several contingency cells are sparse.}
\small

\begin{tabular}[t]{llrrr}
\toprule
Variable & Test & $H$ & df & $p$\\
\midrule
Age & Kruskal--Wallis & 77.89 & 2 & $<$.001\\
Gender & Fisher exact (MC) &  &  & $<$.001\\
Musical experience & Fisher exact (MC) &  &  & $<$.001\\
Food experience & Fisher exact (MC) &  &  & $<$.001\\
Listening device & Fisher exact (MC) &  &  & .091\\
Hearing impairment & Fisher exact (MC) &  &  & .800\\
Taste impairment & Fisher exact (MC) &  &  & .736\\
\bottomrule
\end{tabular}
\end{table}
\noindent\emph{How to read it.} A small $p$ means the three cohorts differ on that variable. Every variable except the two impairment screens differs, which is the justification for carrying gender, musical experience and food experience as covariates in the rating models.

\subsubsection*{Sample-size determination}

The per-cohort target of roughly 80 participants comes from a Monte Carlo power analysis run
before data collection. Closed-form power formulas are not available for this design --- a
repeated-measures structure in which each participant rates twelve descriptors for three of the
four prompts --- so power was estimated by simulation instead.

The procedure was as follows. The main effect of \emph{prompt} in the preliminary Italian sample
($N = 90$) of Spanio et al. supplied an empirical baseline effect size, expressed as partial
$\eta^2$ and converted to Cohen's $f$. Three scenarios were then defined by scaling that
baseline (small, medium, large). For each scenario and each candidate sample size from 20 to 150
participants in steps of 10, 250 synthetic datasets were generated under the same ANOVA
specification used in the analysis, and power was taken as the proportion of simulated datasets
in which the prompt term reached significance. The smallest sample size reaching $80\%$ power
under the medium scenario was adopted as the recruitment target: approximately 80 participants
per country. All three cohorts exceeded it (Japan 140, Argentina 104, Italy 117), so the study is
somewhat better powered than planned for the effect it was designed around.

The full simulation report, including the power curves for all three scenarios and the code that
produces them, is archived with the analysis repository on Zenodo as
\texttt{power\_analysis.pdf}.

\subsection{Task 1: model preference}

Task~1 scores run on a 0--10 scale with 5 as the neutral point; values above 5 favour the
fine-tuned model. Because the five trials are nested within participants, every inferential
test below is run on participant means, and the trial-level structure is handled separately by
the mixed models in Table~\ref{tab:t1lmm}.

\begin{table}[H]
\centering
\caption{Task 1 within-cohort tests against the neutral point of 5 (one-sided Wilcoxon signed-rank on participant mean scores). $r$ is the rank-biserial correlation with a 2,000-resample bootstrap percentile interval.}
\label{tab:t1within}
\small

\begin{tabular}[t]{lrrrrrrr}
\toprule
Group & $N$ & Median & Mean & SD & $r$ & 95\% CI & $p$\\
\midrule
Japan & 140 & 5.00 & 4.85 & 1.53 & -.147 & {}[-.339, .053] & .927\\
Argentina & 104 & 5.60 & 5.42 & 1.70 & .292 & {}[.080, .506] & .006\\
Italy & 117 & 5.20 & 5.47 & 1.59 & .301 & {}[.093, .503] & .003\\
\bottomrule
\end{tabular}
\end{table}
\noindent\emph{How to read it.} $r$ is a non-parametric effect size bounded by $\pm 1$; positive values mean the fine-tuned model was preferred. A confidence interval that excludes 0 is the effect-size counterpart of a significant $p$. Argentina and Italy clear the neutral point, Japan does not.

\begin{table}[H]
\centering
\caption{Task 1 within-cohort tests by taste prompt (one-sided Wilcoxon signed-rank on participant $\times$ prompt means). $p_{\text{adj}}$ is Bonferroni-corrected within cohort across the four prompts.}
\small

\begin{tabular}[t]{llrrrrr}
\toprule
Group & Prompt & $N$ & Median & $r$ & $p$ & $p_{\text{adj}}$\\
\midrule
Japan & Bitter & 106 & 5.00 & -.030 & .600 & 1.000\\
Japan & Salty & 111 & 4.67 & -.354 & .998 & 1.000\\
Japan & Sour & 102 & 5.00 & -.081 & .749 & 1.000\\
Japan & Sweet & 114 & 5.50 & .169 & .068 & .270\\
Argentina & Bitter & 85 & 5.00 & -.201 & .931 & 1.000\\
Argentina & Salty & 81 & 5.00 & -.157 & .873 & 1.000\\
Argentina & Sour & 83 & 7.00 & .445 & $<$.001 & .001\\
Argentina & Sweet & 81 & 7.00 & .533 & $<$.001 & $<$.001\\
Italy & Bitter & 92 & 5.00 & .257 & .021 & .085\\
Italy & Salty & 92 & 4.00 & -.459 & 1.000 & 1.000\\
Italy & Sour & 94 & 6.00 & .344 & .003 & .010\\
Italy & Sweet & 87 & 7.00 & .604 & $<$.001 & $<$.001\\
\bottomrule
\end{tabular}
\end{table}
\noindent\emph{How to read it.} This locates the fine-tuning advantage. It is carried by sweet and sour in Argentina and Italy; salty is at or below the neutral point in all three cohorts, and negative $r$ values there mean listeners preferred the \emph{base} model's salty clips.

\begin{table}[H]
\centering
\caption{Task 1 between-cohort Kruskal--Wallis tests on participant mean scores, pooled and per taste prompt. Per-prompt tests are reported uncorrected.}
\small

\begin{tabular}[t]{lrrrr}
\toprule
Prompt & $H$ & df & $p$ & $\eta^2$\\
\midrule
Overall (pooled) & 11.47 & 2 & .003 & .026\\
Bitter & 4.74 & 2 & .094 & .010\\
Salty & 3.06 & 2 & .216 & .004\\
Sour & 13.37 & 2 & .001 & .041\\
Sweet & 16.45 & 2 & $<$.001 & .052\\
\bottomrule
\end{tabular}
\end{table}
\noindent\emph{How to read it.} This asks whether the three cohorts differ from \emph{each other}, rather than from the neutral point. The country difference is concentrated on sweet and sour; bitter and salty do not separate the cohorts. $\eta^2$ is the $H$-based effect size.

\begin{table}[H]
\centering
\caption{Task 1 trial-level linear mixed models, one per cohort, with random intercepts for participant and stimulus. The estimate is the mean preference above the neutral point of 5. A single cross-cohort model with a fixed effect of country gives $p = .003$ for the country term.}
\label{tab:t1lmm}
\small

\begin{tabular}[t]{lrrr}
\toprule
Group & Estimate & SE & $p$\\
\midrule
Japan & -0.150 & 0.151 & .327\\
Argentina & 0.425 & 0.167 & .012\\
Italy & 0.472 & 0.193 & .021\\
\bottomrule
\end{tabular}
\end{table}
\noindent\emph{How to read it.} These models keep the five trials as separate observations instead of averaging them, so they are a check that the participant-level conclusion is not an artefact of aggregation. The signs, sizes and significance pattern reproduce Table~\ref{tab:t1within}.

\subsection{Task 2: perception-rating models}

The manuscript reports a condensed version of the two omnibus ANOVAs, giving only $df$, $F$, $p$
and partial $\eta^2$. The full tables below add the sums of squares, the intercept and the
residual line. Both models use Type~III sums of squares with sum-to-zero contrasts on all
categorical predictors, and are fitted to respondents reporting Male or Female gender who are
not Professional Eaters.

\begin{table}[H]
\centering
\caption{Full Type III ANOVA of the raw Task 2 ratings.}
\label{tab:anovaraw}
\small

\begin{tabular}[t]{lrrrrr}
\toprule
Term & Sum Sq & df & $F$ & $p$ & partial $\eta^2$\\
\midrule
(Intercept) & 46,355.0 & 1 & 40,023.44 & $<$.001 & .773\\
Group & 1,010.5 & 2 & 436.22 & $<$.001 & .069\\
Prompt & 80.9 & 3 & 23.28 & $<$.001 & .006\\
Descriptor & 377.7 & 11 & 29.64 & $<$.001 & .027\\
Musical experience & 73.3 & 2 & 31.62 & $<$.001 & .005\\
Food experience & 4.1 & 1 & 3.55 & .060 & $<$.001\\
Gender & 1.1 & 1 & 0.91 & .340 & $<$.001\\
Group $\times$ Prompt & 16.9 & 6 & 2.43 & .024 & .001\\
Group $\times$ Descriptor & 367.6 & 22 & 14.43 & $<$.001 & .026\\
Prompt $\times$ Descriptor & 977.8 & 33 & 25.58 & $<$.001 & .067\\
Group $\times$ Prompt $\times$ Descriptor & 226.2 & 66 & 2.96 & $<$.001 & .016\\
Residuals & 13,629.6 & 11,768 &  &  & \\
\bottomrule
\end{tabular}
\end{table}
\noindent\emph{How to read it.} The manuscript's condensed table drops the Sum Sq column, the intercept and the residual line shown here. Partial $\eta^2$ is $SS_{\text{term}} / (SS_{\text{term}} + SS_{\text{residual}})$: the share of variance a term accounts for once the others are held constant.

\begin{table}[H]
\centering
\caption{Full Type III ANOVA of the within-subject z-scored Task 2 ratings. One participant with zero within-subject variance is excluded.}
\label{tab:anovaz}
\small

\begin{tabular}[t]{lrrrrr}
\toprule
Term & Sum Sq & df & $F$ & $p$ & partial $\eta^2$\\
\midrule
(Intercept) & 0.1 & 1 & 0.07 & .792 & $<$.001\\
Group & 0.0 & 2 & 0.00 & .998 & $<$.001\\
Prompt & 29.2 & 3 & 11.51 & $<$.001 & .003\\
Descriptor & 329.5 & 11 & 35.44 & $<$.001 & .032\\
Musical experience & 0.1 & 2 & 0.03 & .969 & $<$.001\\
Food experience & 0.1 & 1 & 0.08 & .779 & $<$.001\\
Gender & 0.0 & 1 & 0.01 & .911 & $<$.001\\
Group $\times$ Prompt & 3.1 & 6 & 0.61 & .723 & $<$.001\\
Group $\times$ Descriptor & 289.0 & 22 & 15.54 & $<$.001 & .028\\
Prompt $\times$ Descriptor & 792.4 & 33 & 28.41 & $<$.001 & .074\\
Group $\times$ Prompt $\times$ Descriptor & 177.1 & 66 & 3.18 & $<$.001 & .018\\
Residuals & 9,915.7 & 11,732 &  &  & \\
\bottomrule
\end{tabular}
\end{table}
\noindent\emph{How to read it.} Compare row by row with Table~\ref{tab:anovaraw}. Normalizing within participant removes each person's own mean and spread, so terms that survive here describe the \emph{shape} of the mapping rather than the level at which a cohort rates.

\begin{table}[H]
\centering
\caption{Per-cohort Type III ANOVAs of the Task 2 ratings, fitted separately within each cohort without the \emph{group} factor. Residual and intercept rows omitted.}
\footnotesize

\begin{tabular}[t]{lrrrrrrrrr}
\toprule
\multicolumn{1}{c}{ } & \multicolumn{3}{c}{Japan} & \multicolumn{3}{c}{Argentina} & \multicolumn{3}{c}{Italy} \\
\cmidrule(l{3pt}r{3pt}){2-4} \cmidrule(l{3pt}r{3pt}){5-7} \cmidrule(l{3pt}r{3pt}){8-10}
Term & $F$  & $p$  & $\eta_p^2$  & $F$   & $p$   & $\eta_p^2$   & $F$    & $p$    & $\eta_p^2$   \\
\midrule
Prompt & 16.59 & $<$.001 & .011 & 4.82 & .002 & .004 & 7.36 & $<$.001 & .006\\
Descriptor & 34.76 & $<$.001 & .077 & 7.92 & $<$.001 & .024 & 16.55 & $<$.001 & .047\\
Musical experience & 7.00 & $<$.001 & .003 & 34.83 & $<$.001 & .020 & 23.36 & $<$.001 & .012\\
Food experience & 2.99 & .084 & .001 & 0.65 & .420 & $<$.001 & 3.11 & .078 & .001\\
Gender & 0.36 & .548 & $<$.001 & 0.15 & .695 & $<$.001 & 1.52 & .217 & $<$.001\\
Prompt $\times$ Descriptor & 14.03 & $<$.001 & .092 & 12.38 & $<$.001 & .105 & 5.96 & $<$.001 & .050\\
Residuals &  &  &  &  &  &  &  &  & \\
\bottomrule
\end{tabular}
\end{table}
\noindent\emph{How to read it.} The Prompt $\times$ Descriptor row measures how strongly the generating prompt differentiates a cohort's rating profile --- but not whether that differentiation follows the \emph{intended} diagonal. Japan and Argentina score highest here even though both cross-map bitter, because a systematic off-diagonal mapping differentiates a profile just as much as a correct one does. Read this row together with the heatmaps: it says how much taste structure a cohort hears, the heatmaps say which structure.

\begin{table}[H]
\centering
\caption{Type III tests from the linear mixed models with crossed random intercepts for participant and stimulus song, fitted to the raw and to the within-subject z-scored ratings (Satterthwaite denominator degrees of freedom).}
\small

\begin{tabular}[t]{lrrrrrr}
\toprule
\multicolumn{3}{c}{ } & \multicolumn{2}{c}{Raw ratings} & \multicolumn{2}{c}{Within-subject $z$} \\
\cmidrule(l{3pt}r{3pt}){4-5} \cmidrule(l{3pt}r{3pt}){6-7}
Term & NumDF & DenDF & $F$  & $p$  & $F$   & $p$  \\
\midrule
Group & 2 & 328 & 92.51 & $<$.001 & 0.00 & .998\\
Prompt & 3 & 9,466 & 18.77 & $<$.001 & 11.67 & $<$.001\\
Descriptor & 11 & 11,559 & 33.63 & $<$.001 & 35.89 & $<$.001\\
Musical experience & 2 & 329 & 6.70 & .001 & 0.03 & .970\\
Food experience & 1 & 330 & 0.83 & .362 & 0.08 & .773\\
Gender & 1 & 329 & 0.29 & .589 & 0.02 & .891\\
Group $\times$ Prompt & 6 & 9,411 & 0.99 & .428 & 0.62 & .714\\
Group $\times$ Descriptor & 22 & 11,559 & 16.37 & $<$.001 & 15.74 & $<$.001\\
Prompt $\times$ Descriptor & 33 & 11,559 & 29.02 & $<$.001 & 28.77 & $<$.001\\
Group $\times$ Prompt $\times$ Descriptor & 66 & 11,559 & 3.36 & $<$.001 & 3.22 & $<$.001\\
\bottomrule
\end{tabular}
\end{table}
\noindent\emph{How to read it.} Each participant contributed 36 nested ratings and several participants rated the same excerpt; these models account for both dependencies explicitly. A term is treated as supported in the manuscript only when the mixed model agrees with the ANOVA. The one disagreement is Group $\times$ Prompt, which is significant in the raw ANOVA but not here.  Random-effect standard deviations, raw model: participant $0.35$, song $0.04$, residual $1.01$; z-scored model: participant $0.00$, song $0.01$, residual $0.91$.  Song-level variance is small relative to the residual, so stimulus idiosyncrasy is controlled for without driving the cross-cultural result.

\begin{table}[H]
\centering
\caption{Group-related Type III $p$-values under the main model and two sensitivity variants. \emph{Retain excluded} adds back the respondents with a gender other than Male or Female and the Professional Eaters; \emph{Exclude legacy IT} drops the legacy Italian collection wave.}
\small

\begin{tabular}[t]{lrrr}
\toprule
Term & Main model & Retain excluded & Exclude legacy IT\\
\midrule
Group & $<$.001 & $<$.001 & $<$.001\\
Group $\times$ Prompt & .024 & .034 & .022\\
Group $\times$ Descriptor & $<$.001 & $<$.001 & $<$.001\\
Group $\times$ Prompt $\times$ Descriptor & $<$.001 & $<$.001 & $<$.001\\
\bottomrule
\end{tabular}
\end{table}
\noindent\emph{How to read it.} A conclusion that holds across all three columns is not an artefact of the sample-construction choices. The legacy wave is 90 of the 117 Italian participants, so the last column rests on only 27 newly collected Italian respondents and is correspondingly low-powered: read it as a check on direction, not as a precise re-estimate.

\begin{table}[H]
\centering
\caption{Italian-only mixed model testing the effect of data-collection wave (legacy vs newly collected) and its interaction with the prompt $\times$ descriptor structure.}
\small

\begin{tabular}[t]{lrrrr}
\toprule
Term & NumDF & DenDF & $F$ & $p$\\
\midrule
Wave & 1 & 116 & 0.06 & .808\\
Wave $\times$ Prompt & 3 & 3,903 & 0.33 & .801\\
Wave $\times$ Descriptor & 11 & 4,001 & 1.27 & .235\\
Wave $\times$ Prompt $\times$ Descriptor & 33 & 4,001 & 1.13 & .282\\
\bottomrule
\end{tabular}
\end{table}
\noindent\emph{How to read it.} Non-significant wave terms mean the two Italian waves are statistically interchangeable for this analysis, which is what justifies pooling them into a single Italian cohort.

\subsection{Response style}

A response style is a habit of using a rating scale in a particular way regardless of item
content. Two indices capture the relevant habits: acquiescence (ARS), a participant's mean
across their 36 ratings, and extreme responding (ERS), the proportion of those ratings at
either end of the scale.

\begin{table}[H]
\centering
\caption{Per-cohort response-style indices. ARS is the participant mean across their 36 ratings, ERS the proportion of those ratings equal to 1 or 5, Within-SD the within-participant standard deviation.}
\label{tab:style}
\small

\begin{tabular}[t]{lrrrrrr}
\toprule
Group & $n$ & ARS mean & ARS SD & ERS mean & ERS SD & Within-SD\\
\midrule
Japan & 128 & 2.70 & 0.36 & 0.26 & 0.19 & 1.17\\
Argentina & 98 & 1.98 & 0.45 & 0.48 & 0.20 & 0.99\\
Italy & 105 & 1.97 & 0.40 & 0.48 & 0.19 & 1.00\\
\bottomrule
\end{tabular}
\end{table}
\noindent\emph{How to read it.} Japanese respondents rate higher on average (ARS) but reach the ends of the scale less often (ERS) than the other two cohorts. That combination inflates a raw country main effect without any difference in what was heard, which is why the manuscript re-fits the model on within-subject z-scores.

\begin{table}[H]
\centering
\caption{Kruskal--Wallis tests of the response-style indices across cohorts.}
\small

\begin{tabular}[t]{lrrr}
\toprule
Index & $H$ & df & $p$\\
\midrule
ARS & 151.11 & 2 & $<$.001\\
ERS & 75.64 & 2 & $<$.001\\
Within-SD & 35.81 & 2 & $<$.001\\
\bottomrule
\end{tabular}
\end{table}
\noindent\emph{How to read it.} All three indices differ across cohorts, confirming that scale use is not comparable between the three groups before normalization.

\begin{table}[H]
\centering
\caption{Levene's tests of variance homogeneity across cohorts, overall and separately for each of the twelve descriptors (median-centred statistic).}
\small

\begin{tabular}[t]{lrrrr}
\toprule
Descriptor & $F$ & df1 & df2 & $p$\\
\midrule
Overall & 136.21 & 2 & 11913 & $<$.001\\
Anger & 18.76 & 2 & 990 & $<$.001\\
Bitter & 0.76 & 2 & 990 & .466\\
Cold & 0.42 & 2 & 990 & .660\\
Disgust & 51.90 & 2 & 990 & $<$.001\\
Fear & 11.59 & 2 & 990 & $<$.001\\
Happy & 5.56 & 2 & 990 & .004\\
Hot & 3.32 & 2 & 990 & .037\\
Sad & 0.36 & 2 & 990 & .695\\
Salty & 15.15 & 2 & 990 & $<$.001\\
Sour & 6.75 & 2 & 990 & .001\\
Surprise & 0.32 & 2 & 990 & .729\\
Sweet & 4.52 & 2 & 990 & .011\\
\bottomrule
\end{tabular}
\end{table}
\noindent\emph{How to read it.} These check the equal-variance assumption of the ANOVA. Heteroscedasticity is present but tolerable at this sample size; the mixed models and the z-scored re-fit are the substantive safeguards against it.

\begin{figure}[H]
\centering
\includegraphics[width=\linewidth]{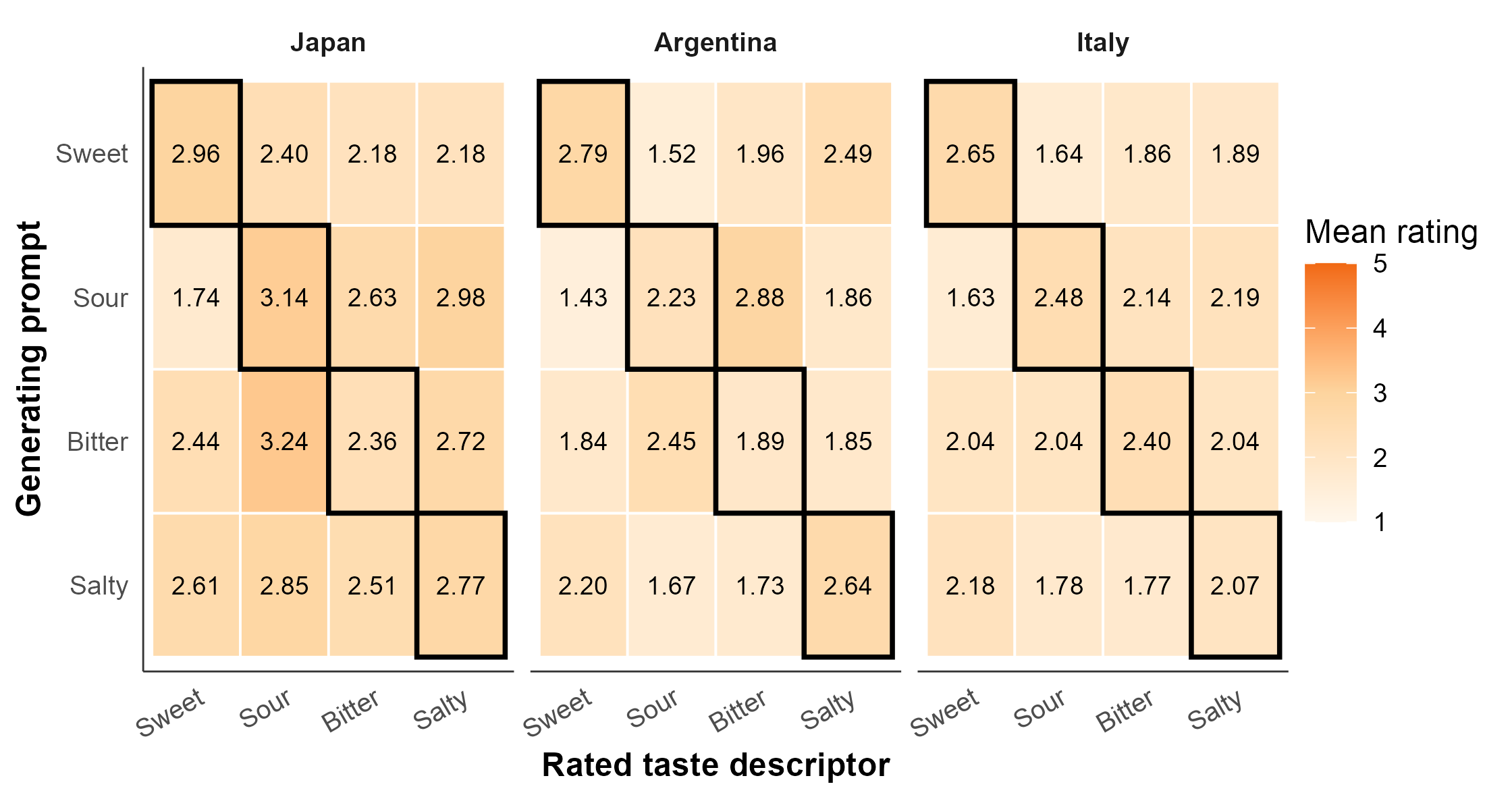}
\caption{Mean \emph{raw} taste ratings by cohort. Rows are the prompt used to generate the clip, columns the taste descriptor rated; black outlines mark the matching prompt--descriptor cells.}
\end{figure}
\noindent\emph{How to read it.} This is the uncorrected counterpart of the z-scored heatmap in the manuscript. The whole Japanese panel sits at a higher level than the other two, which is the response-style offset; the \emph{arrangement} of high and low cells within each panel is the part that survives normalization.

\begin{figure}[H]
\centering
\includegraphics[width=0.92\linewidth]{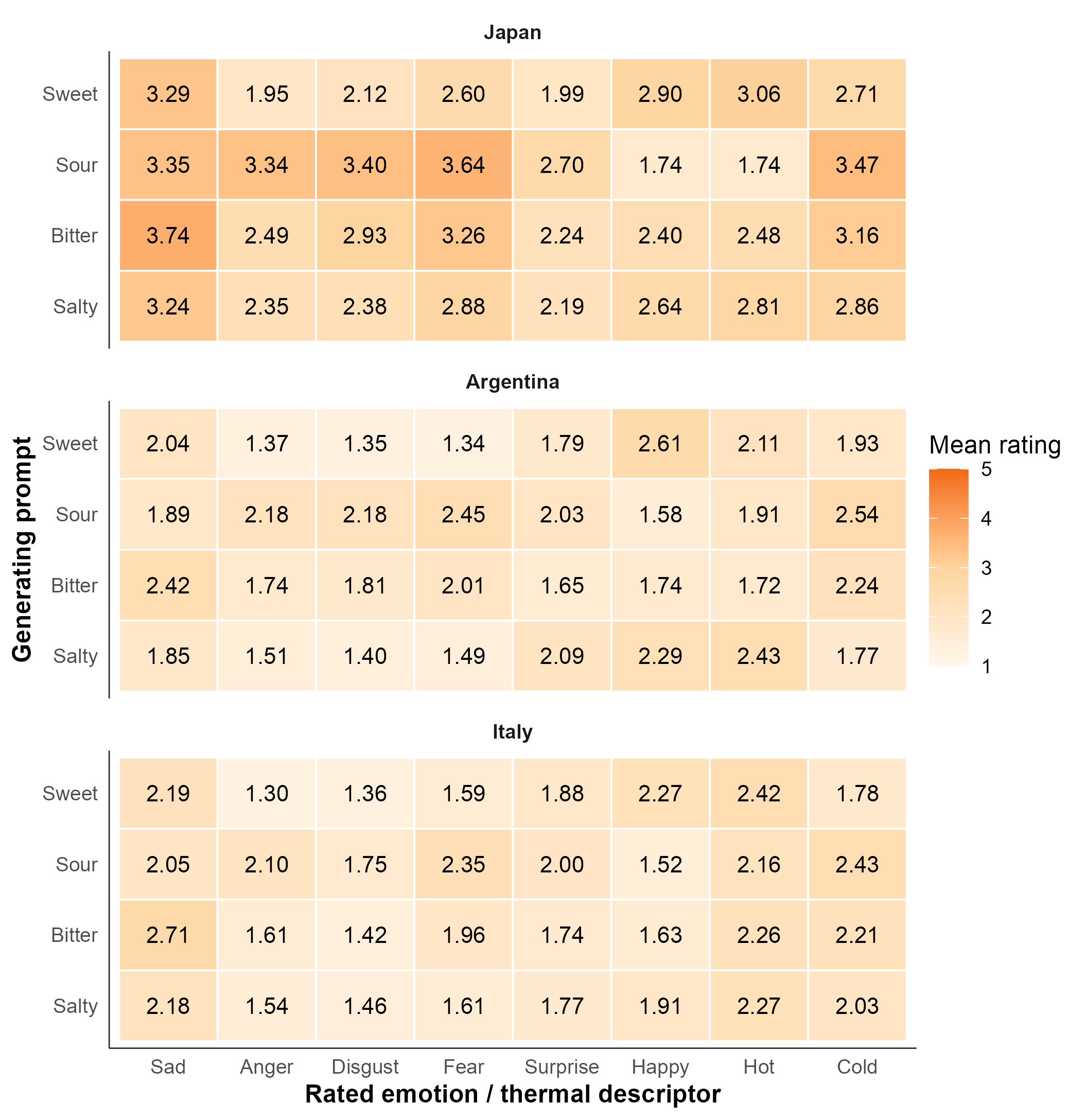}
\caption{Mean \emph{raw} emotion and thermal ratings by cohort.}
\end{figure}
\noindent\emph{How to read it.} Sweet clips are rated happier and warmer, sour and bitter clips colder and more negative, most clearly in Argentina and Italy. The Japanese panel is uniformly elevated across both positive and negative descriptors, the same acquiescence pattern seen in Table~\ref{tab:style}.

\begin{figure}[H]
\centering
\includegraphics[width=0.92\linewidth]{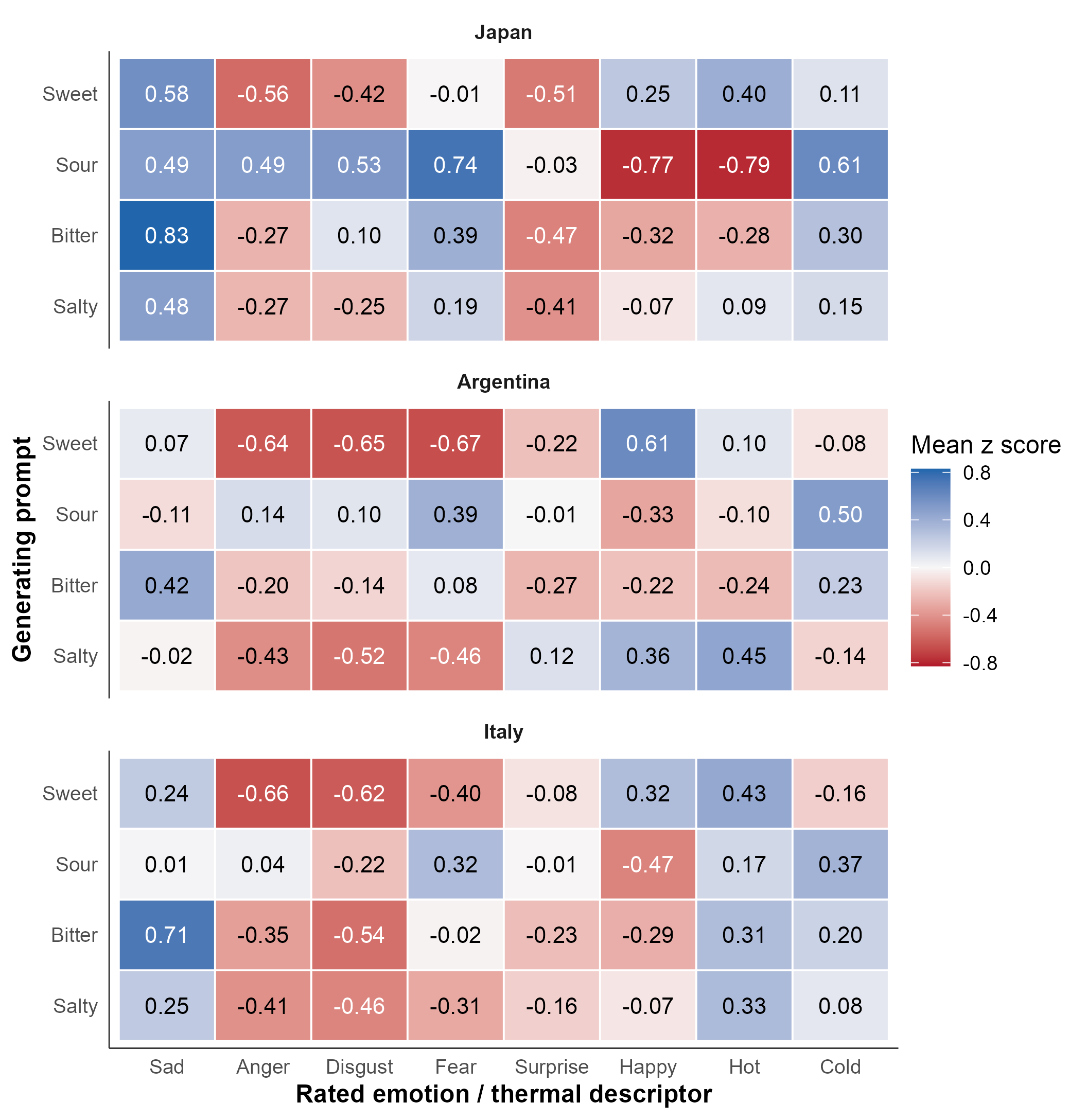}
\caption{Mean within-subject z-scored emotion and thermal ratings by cohort.}
\end{figure}
\noindent\emph{How to read it.} With the level difference removed, a structural difference remains: the sour prompt recruits negative-affect descriptors most strongly in Japan, the bitter prompt aligns chiefly with sadness in Japan and Italy, and the sweet prompt aligns with happy and hot everywhere but to different degrees.

\subsection{Inter-annotator agreement}

The heatmaps average over listeners, so a cohort could show a clean mean structure while its
individual members disagreed completely. Treating the raters of each (song, taste descriptor)
cell as annotators quantifies how much of the structure is shared within a cohort.

\begin{table}[H]
\centering
\caption{Inter-annotator agreement on the four taste descriptors by cohort. The first five columns are mean pairwise distances on the 1--5 scale rescaled to $[0, 1]$ (smaller is better); the last two are Krippendorff's $\alpha$ aggregated over the four taste descriptors.}
\small

\begin{tabular}[t]{lrrrrrrr}
\toprule
Group & Salty & Sweet & Bitter & Sour & Mean & $\alpha$ (interval) & $\alpha$ (ordinal)\\
\midrule
Japan & .281 & .279 & .281 & .314 & .289 & .146 & .143\\
Argentina & .276 & .241 & .286 & .282 & .271 & .191 & .190\\
Italy & .256 & .265 & .283 & .264 & .267 & .099 & .099\\
\bottomrule
\end{tabular}
\end{table}
\noindent\emph{How to read it.} Mean pairwise distance is the average absolute gap between two raters of the same cell: $0$ is perfect agreement. Krippendorff's $\alpha$ is $1$ for perfect agreement and $0$ for chance. Agreement is low in every cohort, as expected for single-trial perceptual ratings of AI-generated audio by untrained listeners, and lowest in Italy, whose bitter cell falls to chance. The interval and ordinal metrics agree to within $\pm 0.01$.

The song-by-descriptor breakdown behind this table, for all twelve descriptors and for the
pooled sample as well as the three cohorts, is provided as \textbf{S1 File}.

\subsection{Factor structure}

Exploratory factor analysis groups descriptors that move together into a smaller number of
latent dimensions. The manuscript shows the three cohort-specific solutions; this section adds
the pooled solution and the stability check behind the congruence coefficients.

\begin{figure}[H]
\centering
\includegraphics[width=0.82\linewidth]{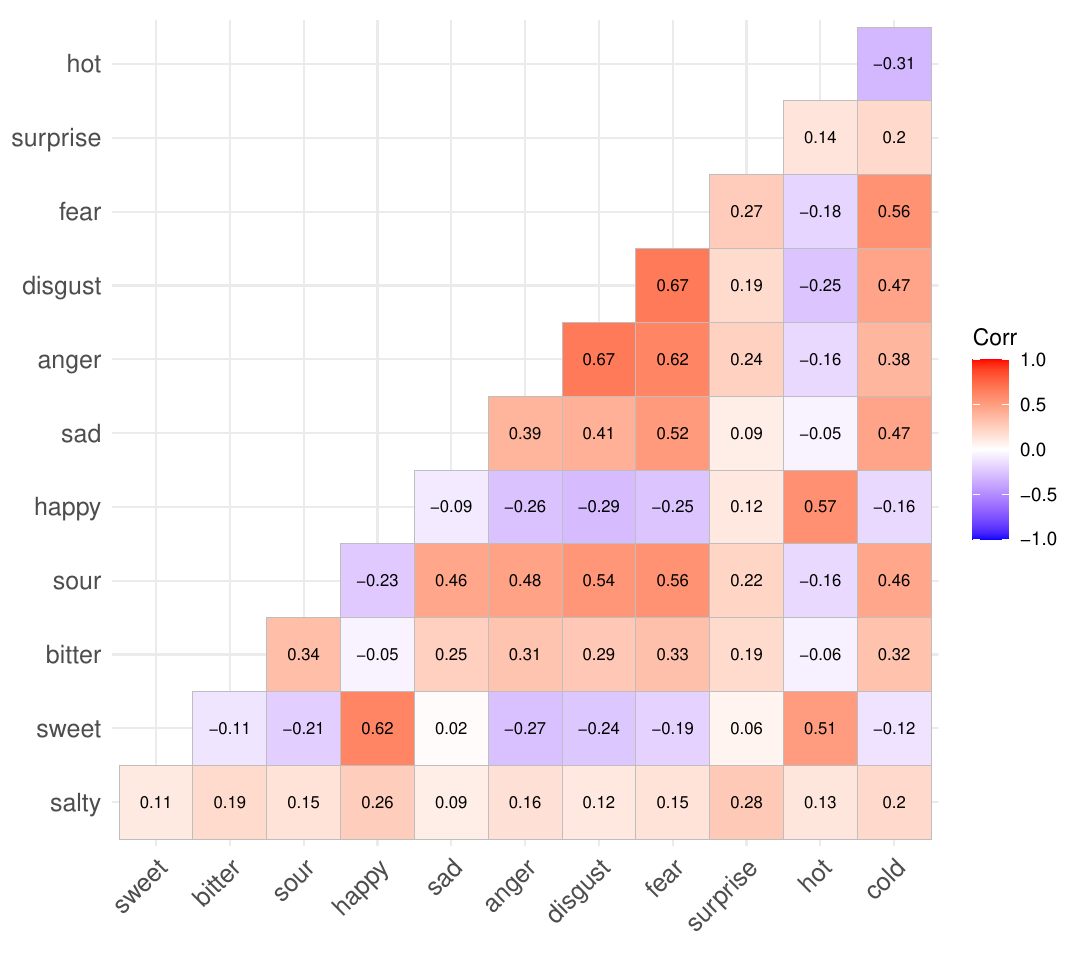}
\caption{Correlation matrix of the twelve descriptors in the pooled three-cohort sample.}
\end{figure}
\noindent\emph{How to read it.} Blocks of mutually correlated descriptors are what the factor analysis formalizes. The negative-affect descriptors form the most visible block; sweet, happy and hot form a second.

\begin{figure}[H]
\centering
\includegraphics[width=0.82\linewidth]{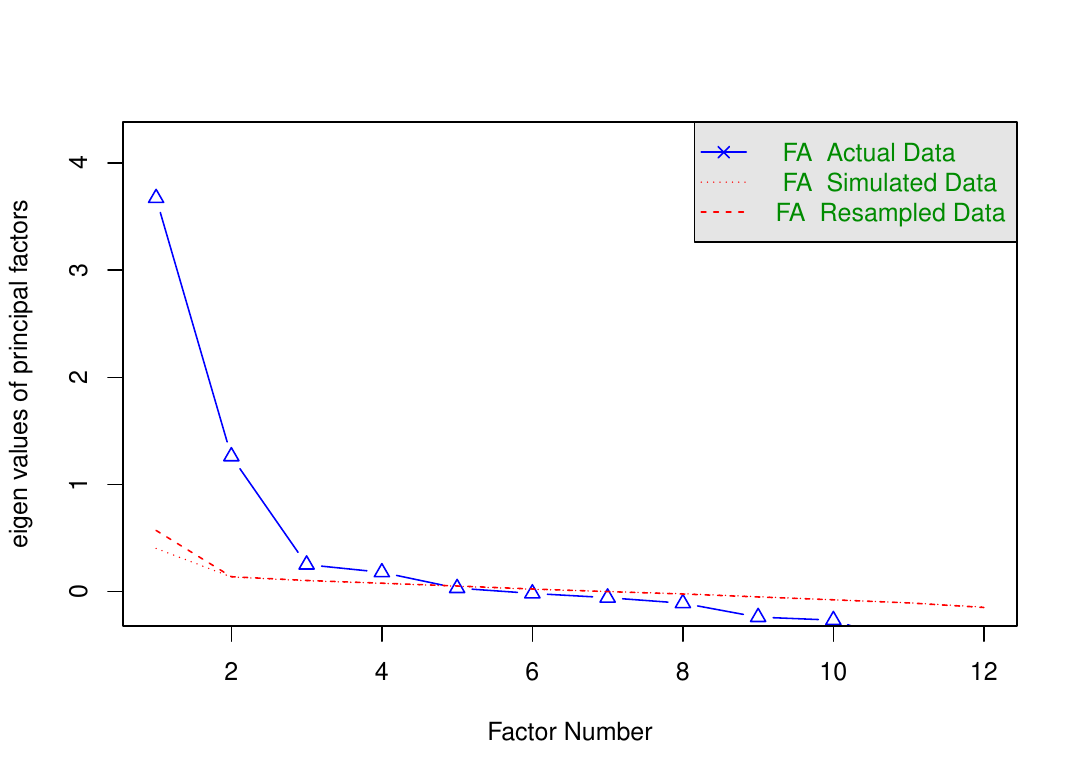}
\caption{Parallel-analysis scree plot for the pooled sample. The analysis retains 4 factors.}
\end{figure}
\noindent\emph{How to read it.} Parallel analysis compares the eigenvalues of the real data with those of random data of the same size; factors are retained while the real eigenvalue exceeds the simulated one. This is the rule used to choose the number of factors for each cohort separately.

\begin{figure}[H]
\centering
\includegraphics[width=0.82\linewidth]{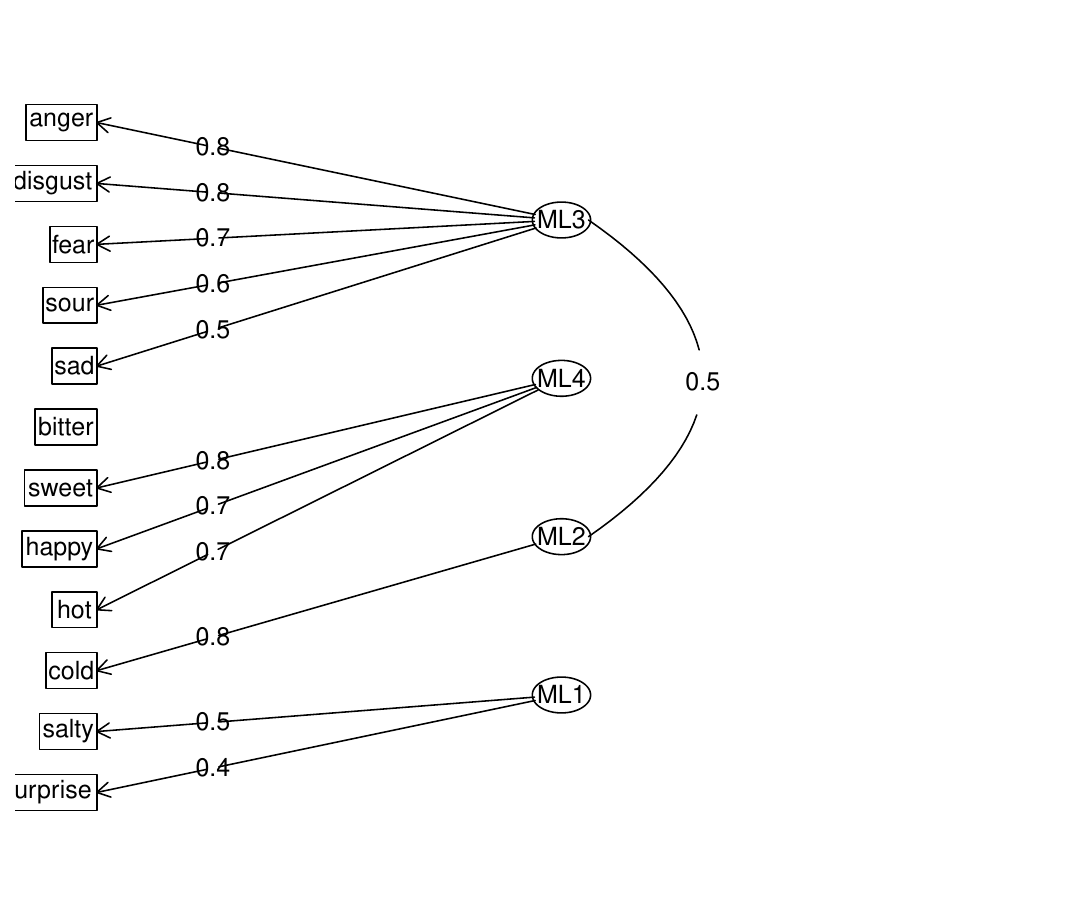}
\caption{Pooled factor solution (4 factors, maximum-likelihood extraction, oblimin rotation).}
\end{figure}
\noindent\emph{How to read it.} The pooled solution is reported for completeness only. Because the three cohorts organize the descriptor space differently, pooling them averages over exactly the difference the study is about; the cohort-specific solutions in the manuscript are the informative ones.

\begin{table}[H]
\centering
\caption{Tucker's congruence coefficients between the three cohort-specific factor solutions (truncated to 3 factors for comparability) with a participant bootstrap. Values above $0.95$ are conventionally read as factor equivalence, values in $[0.85, 0.95)$ as fair similarity.}
\small

\begin{tabular}[t]{lrrrr}
\toprule
Cohort pair & Congruence & Bootstrap median & 95\% interval & $B$\\
\midrule
Japan -- Argentina & .613 & .717 & {}[.635, .902] & 300\\
Japan -- Italy & .777 & .757 & {}[.685, .875] & 300\\
Argentina -- Italy & .460 & .782 & {}[.697, .943] & 300\\
\bottomrule
\end{tabular}
\end{table}
\noindent\emph{How to read it.} The point congruence is the mean of the diagonal of the cross-solution congruence matrix. The bootstrap statistic instead takes the best match per factor, because factor order and sign are not fixed across resamples; it is therefore slightly more generous by construction. Even so, every upper bound stays below $0.95$, so the low congruence is a stable feature of the data and not a small-sample artefact.

Parallel analysis retains 4 factors for Japan, 3 for Argentina and 4 for Italy; the comparison above truncates all three to 3. The numeric loadings behind the manuscript's loading figure are archived as \texttt{tables/O2\_factor\_loadings.csv}.

\subsection{Acoustic characterization of the stimuli}

The remaining tables describe the 100 fine-tuned clips themselves rather than the listeners.
Following Deng et al., the descriptors are read as relative patterns across prompts rather than
as absolute thresholds.

\begin{table}[H]
\centering
\caption{Acoustic descriptors of the 100 fine-tuned clips by generating prompt (mean $\pm$ SD, 25 clips per prompt).}
\small

\begin{tabular}[t]{lrrrr}
\toprule
Descriptor & Sweet & Sour & Bitter & Salty\\
\midrule
Spectral centroid (Hz) & 448 $\pm$ 90 & 741 $\pm$ 297 & 482 $\pm$ 119 & 393 $\pm$ 136\\
Bandwidth (Hz) & 683 $\pm$ 158 & 1,229 $\pm$ 519 & 741 $\pm$ 205 & 826 $\pm$ 353\\
Rolloff (Hz) & 618 $\pm$ 116 & 1,319 $\pm$ 639 & 727 $\pm$ 243 & 574 $\pm$ 185\\
Fundamental frequency (Hz) & 112 $\pm$ 29 & 250 $\pm$ 389 & 127 $\pm$ 49 & 81 $\pm$ 18\\
Spectral flux (roughness) & 23.1 $\pm$ 4.1 & 40.1 $\pm$ 12.7 & 23.3 $\pm$ 7.8 & 29.9 $\pm$ 8.0\\
Sensory dissonance & 0.025 $\pm$ 0.005 & 0.042 $\pm$ 0.015 & 0.028 $\pm$ 0.009 & 0.039 $\pm$ 0.015\\
Onset strength (articulation) & 0.70 $\pm$ 0.17 & 1.01 $\pm$ 0.26 & 0.63 $\pm$ 0.11 & 0.80 $\pm$ 0.25\\
Tempo (BPM) & 120.1 $\pm$ 21.2 & 121.8 $\pm$ 22.7 & 124.8 $\pm$ 16.4 & 122.2 $\pm$ 18.0\\
Zero-crossing rate & 0.022 $\pm$ 0.006 & 0.024 $\pm$ 0.016 & 0.020 $\pm$ 0.007 & 0.011 $\pm$ 0.006\\
\bottomrule
\end{tabular}
\end{table}
\noindent\emph{How to read it.} Read across a row to see which prompt stands out on that descriptor. Sour is the only prompt with a distinctive signature: brighter, sharper, higher-pitched and rougher than the rest. Salty has the lowest centroid and pitch of all four, the opposite of the crisp, percussive profile that designed salty sounds usually carry.

\begin{table}[H]
\centering
\caption{Pearson correlations, across the 100 clips, between two acoustic cues and the pooled response-style-corrected taste ratings. Cells give $r$ with $p$ in parentheses.}
\small

\begin{tabular}[t]{lrrrr}
\toprule
Cue & Sweet & Sour & Bitter & Salty\\
\midrule
Sensory dissonance & -.47 ($<$.001) & .25 (.012) & .30 (.003) & .18 (.073)\\
Spectral flux (roughness) & -.56 ($<$.001) & .31 (.002) & .41 ($<$.001) & .21 (.035)\\
\bottomrule
\end{tabular}
\end{table}
\noindent\emph{How to read it.} Both cues separate pleasant from unpleasant taste attributions --- strongly negative for sweet, positive for sour and bitter --- but neither separates sour from bitter. That single shared axis is the acoustic substrate a sour--bitter cross-mapping requires.

\begin{table}[H]
\centering
\caption{Within-cohort Pearson correlations between clip-level spectral roughness and the perceived taste ratings, with perceived taste derived separately within each cohort. Cells give $r$ with $p$ in parentheses.}
\small

\begin{tabular}[t]{lrrrr}
\toprule
Group & Clips & Sour & Bitter & Sweet\\
\midrule
Japan & 99 & .30 (.002) & .29 (.004) & -.46 ($<$.001)\\
Argentina & 91 & .07 (.495) & .40 ($<$.001) & -.42 ($<$.001)\\
Italy & 96 & .36 ($<$.001) & .07 (.527) & -.51 ($<$.001)\\
\bottomrule
\end{tabular}
\end{table}
\noindent\emph{How to read it.} This is the clearest single expression of the cross-cultural result. The rough, dissonant region of the stimulus space tracks sourness in Italy, bitterness in Argentina, and both in Japan --- the same sounds, different taste descriptors. Sweetness behaves identically everywhere, so the divergence is specific to the two negative tastes.

\begin{figure}[H]
\centering
\includegraphics[width=\linewidth]{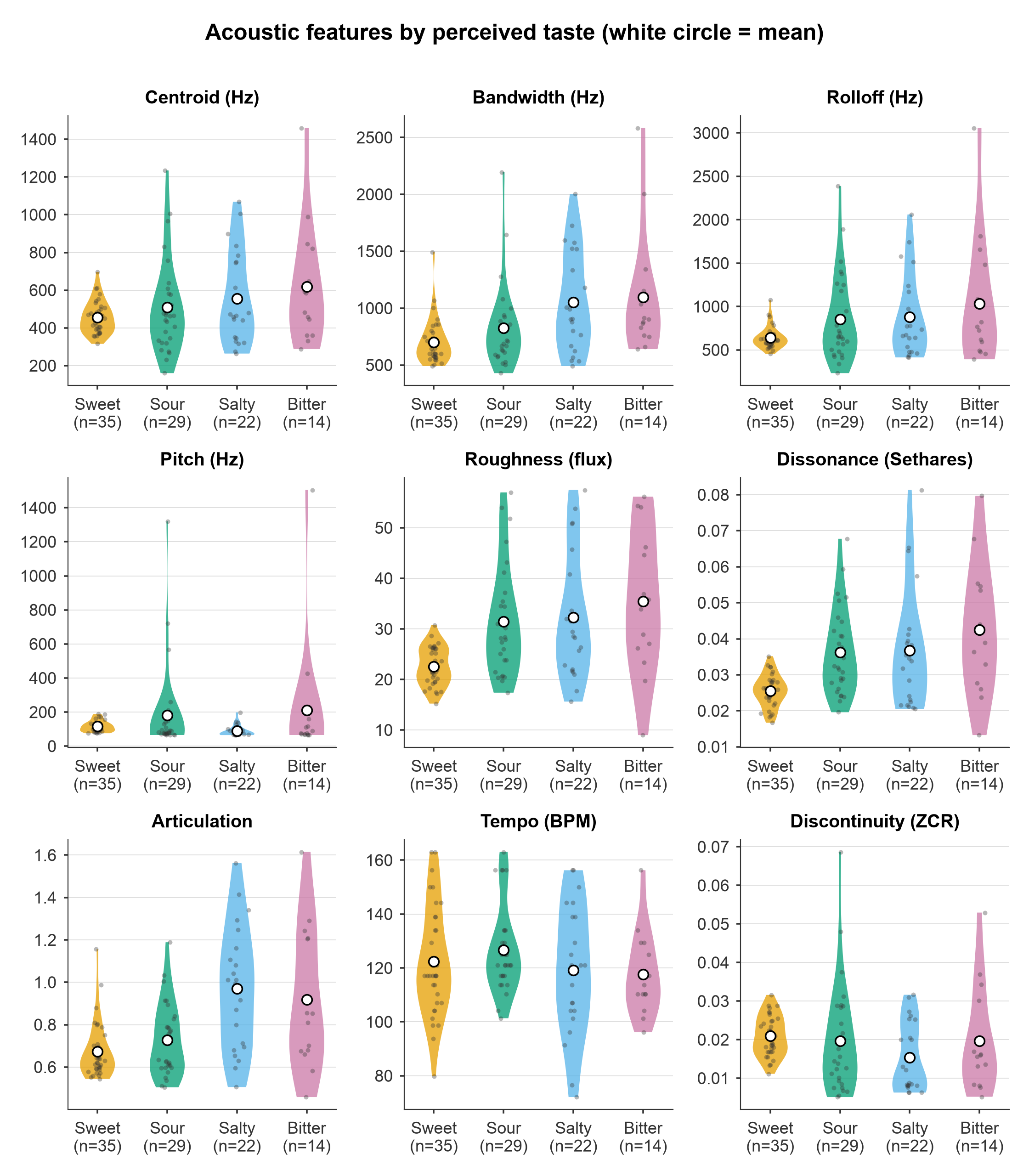}
\caption{Acoustic descriptors of the 100 clips grouped by \emph{perceived} dominant taste pooled over cohorts (the highest of each clip's four response-style-corrected taste ratings).}
\end{figure}
\noindent\emph{How to read it.} Compare with the per-prompt figure in the manuscript: grouping by what listeners heard rather than by what the model was asked for produces cleaner acoustic contrasts. Clips heard as sweet are the softest and most consonant, clips heard as sour or bitter the roughest, and clips heard as salty the most articulated.

\subsection{Software environment and reproducibility}

\begin{table}[H]
\centering
\caption{Software versions used to produce this appendix.}
\small

\begin{tabular}[t]{ll}
\toprule
Component & Version\\
\midrule
R & 4.6.1\\
tidyverse & 2.0.0\\
psych & 2.6.5\\
GPArotation & 2026.8.1\\
car & 3.1.5\\
broom & 1.0.13\\
lme4 & 2.0.6\\
lmerTest & 3.2.1\\
kableExtra & 1.4.1\\
\bottomrule
\end{tabular}
\end{table}
\noindent\emph{How to read it.} All stochastic steps (Fisher Monte Carlo tests, bootstrap confidence intervals, parallel analysis, the congruence bootstrap) are seeded with 20260428, so re-running \texttt{scripts/make\_supplementary.R} reproduces every number above exactly. Exact package versions are pinned in \texttt{renv.lock} in the code repository.

\end{document}